\documentclass{osa-article}

\journal{osajournal}

\articletype{Research Article}
\usepackage{multirow}
\usepackage{tabularx}
\usepackage{caption}

\usepackage{graphicx}
\usepackage{xcolor}

\usepackage{subfig}
\newcommand{\redc}[1]{\textcolor{black}{#1}}

\begin{document}

\title{Direct Estimation of Pupil Parameters Using Deep Learning
for Visible Light Pupillometry}

\author{Abhijeet Phatak,\authormark{1,†} Aditya Chandra Mandal,\authormark{2,†} Janarthanam Jothi Balaji\authormark{3} and Vasudevan Lakshminarayanan\authormark{4,*}}

\address{\authormark{1}598 San Posadas Ter, Sunnyvale, California 94085, USA\\
\authormark{2}Department of Mining Engineering, Indian Institute of Technology (BHU), Varanasi - 221005, India\\
\authormark{3}Department of Optometry, Medical Research Foundation, Chennai - 600 006, India\\
\authormark{4}Theoretical and Experimental Epistemology Lab, School of Optometry and Vision Science, University of Waterloo, Waterloo, Ontario N2L 3G1, Canada\\
\authormark{*}Corresponding author: vengulak@uwaterloo.ca\\
\authormark{†}\textit{These authors contributed equally to this work.}}




\begin{abstract}
Pupil reflex to variations in illumination and associated dynamics are of importance in neurology and ophthalmology. This is typically measured using a near Infrared (IR) pupillometer to avoid Purkinje reflections that appear when strong Visible Light (VL) illumination is present. Previously we demonstrated the use of \redc{deep learning} techniques to accurately detect the pupil pixels (segmentation binary mask) in case of VL images for performing VL pupillometry. \redc{Here, we present a method to obtain the parameters of the elliptical pupil boundary along with the segmentation binary pupil mask}. This eliminates the need for an additional, computationally expensive post-processing step of ellipse fitting and also improves segmentation accuracy. Using the time-varying ellipse parameters of pupil, we can compute the dynamics of the Pupillary Light Reflex (PLR). We also present preliminary evaluations of our deep learning algorithms on experimental data. This work is a significant push in our goal to develop and validate a VL pupillometer based on a smartphone that can be used in the field.
\end{abstract}

\section{INTRODUCTION}
Pupillometry is an important non-invasive, diagnostic tool that can be used to study the measurement of pupil size fluctuations over time in response to light, which is also referred to as Pupillary Light Reflex (PLR) in clinical settings. This method is particularly useful in examining various neurological and ophthalmological disorders, optic nerve injury, head trauma resulting from vehicular accidents, oculomotor nerve damage, brain stem lesions including tumors, as well as the use of central nervous system depressants like barbiturates \cite{hall2018eyeing,Belliveau2021-mo,mirjalili2019value}. Although non-light-related fluctuations in pupil size, which are often considered a sign of arousal via Locus Coeruleus (LC) activity, can be utilized to assess brain state in different species, pupil size is mainly determined by light intensity \cite{reimer2016pupil,lee2016pupil, mazziotti2021meye}.  Because of its utility pupillometry is a very important tool for measuring retinal function in individuals with acquired \cite{feigl2012post,ferrari2010using,karavanaki1994pupil,park2017pupillary,park2016pupillary,sabeti2015multifocal,smith1986simple} and hereditary\cite{collison2015full,kawasaki2014pupil,park2011toward} ocular dysfunction. Clinicians, typically, use a penlight to perform qualitative pupil examination. 
Automated infrared (IR) pupillometers have become available that provide quantitative measures of pupil size and reactivity, allowing assessment in both bright and dark-environments, as well as monitoring pupil dynamics over time, however they are often expensive, not portable, and require trained operators. Visible Light (VL) pupillometry has advantages over the conventional near-IR based pupillometry as it does not require specialized equipment and can provide a better representation of the pupil response in a realistic setting as compared to near-IR illumination. However, in VL, images of the eye exhibit more corneal Fresnel reflections at the \redc{air-cornea} interface (Purkinje image) \cite{mandal2021deep}. Several studies have been conducted recently\cite{morita2016pupil,ricciuti2021pupil,sari2016study,takano2019statistical,tachi2018effective}  where conventional image processing algorithms were applied to VL pupillometry. Using images captured in VL, Morita et al. \cite{morita2016pupil} developed a pupil detection algorithm that first identifies iris edges and uses the Hough transform to find the pupil boundary. However, their method relies on a specific image \redc{acquisition} technique which avoids corneal reflections.
Kitazumi and Nakazawa\cite{kitazumi2018robust} presented a Convolutional Neural Network (CNN) based on U-Net\cite{ronneberger2015u} to carry out pupil segmentation and localization of the pupillary center in VL images for gaze tracking and other applications. Previously, we demonstrated the use of U-Net architecture in segmenting pixels representing the pupil in VL images of human eyes for performing VL pupillometry\cite{mandal2021deep}. Pupillography has shown to be useful in many clinical applications\cite{wilhelm2003clinical}. To study the PLR, extraction of ellipse parameters of the best elliptical contour of the identified pupil pixels must be carried out. An additional post-processing step of using a least squares algorithm \cite{fitzgibbon1996direct}  or other computationally intensive algorithm is carried out to extract the ellipse parameters from \redc{the binary pupil mask also known as the segmentation mask represents the pupil pixels. We refer to the binary pupil mask simply as the pupil mask hereafter. \\In this paper, we further extend our prior work\cite{mandal2021deep} to obtain the pupil mask and extract the ellipse parameters that represents the elliptical pupil boundary}. This obviates the additional step of ellipse fitting and also results in training a better performing neural network model in terms of accuracy. This is a milestone in our effort to build a smartphone based VL pupillometer that uses CNN, utilizing on-device computation, without requiring a processing server or internet access. This is critical in areas with poor access to healthcare or in the field where conventional pupillometers cannot be utilized.

\section{THEORY}
\subsection{Dataset and Data Preprocessing}
\begin{figure}[!ht]
  \centering
  \begin{tabular}[b]{c}
    \includegraphics[width=.10\linewidth]{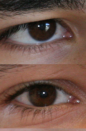} \\
    \small (a)
  \end{tabular} \enspace
  \begin{tabular}[b]{c}
    \includegraphics[width=.10\linewidth]{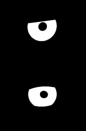} \\
    \small (b)
  \end{tabular} \enspace
   \begin{tabular}[b]{c}
    \includegraphics[width=.10\linewidth]{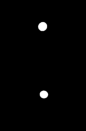} \\
    \small (c)
  \end{tabular} \enspace
  \begin{tabular}[b]{c}
    \includegraphics[width=.10\linewidth]{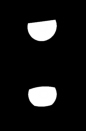} \\
    \small (d)
  \end{tabular} \enspace
  \begin{tabular}[b]{c}
    \includegraphics[width=.10\linewidth]{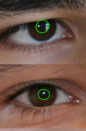} \\
    \small (e)
  \end{tabular}
 \caption[Image dataset] 
   { \label{fig:dataslice} 
Examples of (a) RGB Images from UBIRISv2 (b) Annotated Binary Masks by Hofbauer et.al (c) Extracted Pupil Mask (d) Extracted Binary Iris Mask (e) Ellipse (green) fitted on Pupil in the RGB Image.}                                                                             
\end{figure}

We followed the approach demonstrated in our prior work\cite{mandal2021deep} which was inspired by Kitazumi et.al.\cite{kitazumi2018robust} to compile the initial dataset. The first step in VL pupillometry is identifying the pixels of the pupil in a VL image. For supervised machine learning, we need ground truth output, \textit{i.e.} segmentation mask of the pupil, corresponding to each input image of the human eye. We used the UBIRISv2 \redc{dataset \cite{proencca2009ubiris}} for input images of human eyes in VL. The ground truth pupil masks were obtained from the dataset prepared by Hofbauer et.al.\cite{hofbauer2014ground,alonso2013evaluation} which contained 2250 annotations for a subset of UBIRISv2 (11,102 RGB eye images). The RGB eye images and annotated pupil masks both had width and height of 400 pixels and 300 pixels respectively. Fig.~\ref{fig:dataslice} illustrates some examples from the dataset.


\begin{figure} [!ht]
   \begin{center}
   \begin{tabular}{c} 
   \includegraphics[height=4cm]{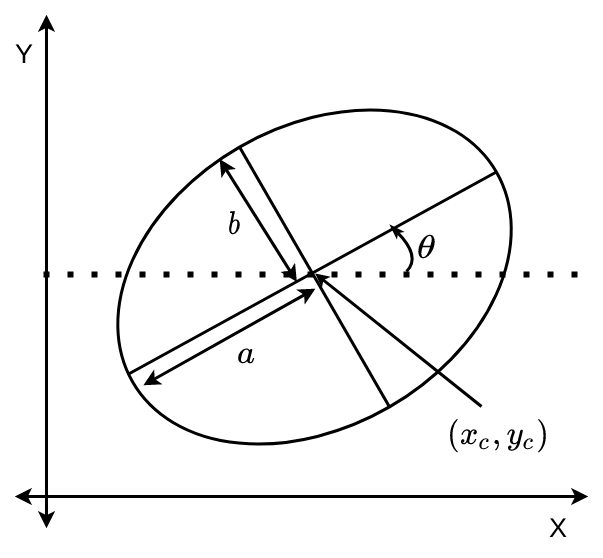}
   \end{tabular}
   \end{center}
   \caption[Definition of an arbitrary ellipse] 
   { \label{fig:ellparam} 
Definition of an elliptical pupil boundary in terms of the co-ordinates of the center $(x_c,y_c)$, semi-major and semi-minor axis lengths $(a,b)$ and the angle of rotation ($\theta$).}
\end{figure}

The most crucial step in VL pupillometry is approximating the pupil contour with an ellipse\cite{wyatt1995form} as we want to observe the variation in the average pupil diameter. An ellipse can be defined by a quintuple - $\varepsilon:(x_c,y_c,2a,2b,\theta)$ where $(x_c,y_c)$ are the coordinates of the center of the pupil. $(a,b)$ are the lengths of the semi-major and semi-minor axis respectively and $\theta$ is the angle of the ellipse measured counterclockwise from the horizontal axis as illustrated in Fig.~\ref{fig:ellparam}.

The ellipse parameters for the pupil masks were obtained using a least squares\cite{fitzgibbon1996direct}  algorithm available in OpenCV. Outliers were removed by ignoring images that had an elliptical pupil contour having a solidity or aspect ratio less than 0.5. \redc{Solidity is the ratio of the area of the elliptical contour to the area of its convex hull. Aspect ratio is defined as ratio of width to height of the bounding rectangle of the ellipse}. The detailed description of data preprocessing step is given in reference\cite{mandal2021deep}.

\subsection{Data Augmentation}

\begin{figure}[!ht]
  \centering
  \begin{tabular}[b]{c}
    \includegraphics[width=.12\linewidth]{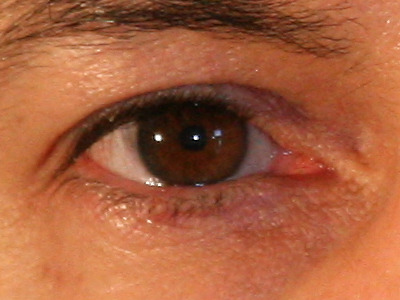} \\
    \small (a)
  \end{tabular} \enspace
  \begin{tabular}[b]{c}
    \includegraphics[width=.12\linewidth]{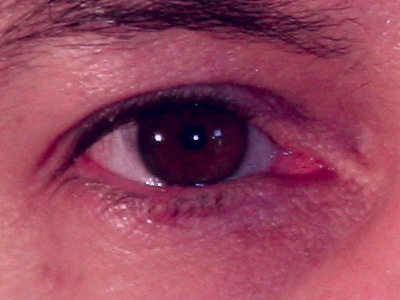} \\
    \small (b)
  \end{tabular} \enspace
   \begin{tabular}[b]{c}
    \includegraphics[width=.12\linewidth]{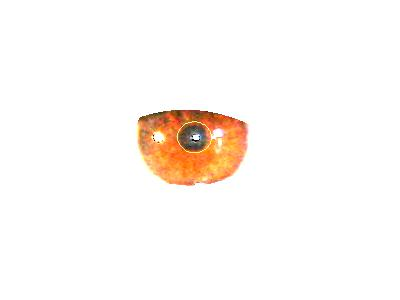} \\
    \small (c)
  \end{tabular} \enspace
  \begin{tabular}[b]{c}
    \includegraphics[width=.12\linewidth]{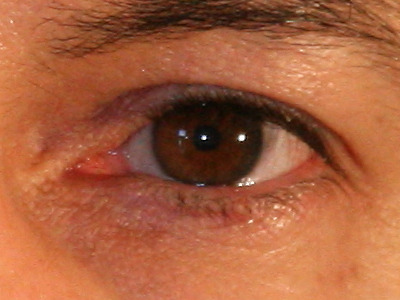} \\
    \small (d)
  \end{tabular} 
\caption[Data Augmentation] 
   { \label{fig:augment} 
Examples of data augmentations performed (a) Original RGB Image from UBIRISv2 (b) Color Augmentation (c) Corneal Reflection Augmentation (d) Flip, Rotate Augmentation
}
\end{figure}

Data Augmentation generates additional data and is important for training neural network models that are robust to noise. The goal is to obtain a model that generalizes well on test (unseen) datasets and does not overfit on the training (seen) dataset. In our prior work\cite{mandal2021deep}, we applied three types of augmentations to the dataset, which are Color Augmentation (CA), Corneal Reflection Augmentation (CRA), and Flip/Rotate Augmentation (FR). Below is a brief description of the augmentations as illustrated in Fig.~\ref{fig:augment}.\\
$\text{CA}$ was necessary to create darker iris and skin tones as compared to the distribution in the dataset (UBIRISv2). CA was applied to the input RGB image (Fig.~\ref{fig:augment}(a)) and no changes were needed for the pupil mask. First the image was converted to Lab color space and noise was added as $[L', a', b'] = [L + rand(L_d), a_i + rand(a_d), b_i + rand(b_d)]$ \redc{where $rand(x)$ refers to a sample drawn uniformly from the continuous range of $x$}, $ -50 < L_d < 10, -20 < a_d < 20, -20 < b_d < 20 $ and $[L,a_i,b_i] = \text{RGB2Lab}([r,g,b])$. Then \redc{$\text{CA}([r, g, b]) = \text{Lab2RGB}([L', a', b'])$}. This helps in dealing with chromatic sensitivity variations and white balance parameters of the camera. An example of CA is shown in Fig.~\ref{fig:augment}(b).\\
CRA was performed to address the corneal reflections in the eye images as VL images show greater corneal (Fresnel) reflections at the \redc{air-cornea} interface when compared to near-IR illumination. This results in eye images with Purkinje images. In VL pupillometry it is expected that the subjects will be looking at intense sources of illumination such as a flashlight or a lamp leading to reflections and saturation. We captured five such images (detailed description is given in reference\cite{mandal2021deep}) referred to as scene images $(I_{sc})$ taken from the subject's perspective. CRA \redc{for an input RGB image $I$} was thus performed using $\text{CRA}(I) = \alpha  I + (1 - \alpha)  I_{iris}  I_{sc}$ where $0.98 < \alpha < 1$. $I_{sc}, I_{iris}$ stand for a scene image and the iris mask respectively. An example of CRA is shown in Fig.~\ref{fig:augment}(c).
Finally, FR was used to improve the rotation invariance of the model by flipping and rotating the images and adjusting the pupil masks accordingly. This was done using standard image processing techniques, and an example of flip augmentation is shown in Fig.~\ref{fig:augment}(d).
\subsection{Deep Learning}
 U-Net has proven highly successful for biomedical image segmentation tasks. It consists of an encoder-decoder architecture that uses multi-stage downsampling with non-linear activations to constrict the representation of the input image, to a latent representation called bottleneck, that is subsequently upsampled by the multi-stage decoder which is expected to output the desired segmentation mask. There are shortcut connections between encoder blocks and the corresponding decoder blocks that pass information. 

\subsubsection{U-Net Model}
We applied the backbone U-Net, as shown in Fig.~\ref{fig:encdec}(a), for image segmentation. We used the Dice Score Coefficient (DSC) as the evaluation criteria. DSC between binary sets $X$, $Y$ is defined in Equation~(\ref{eq:dsc})
\begin{equation}
\label{eq:dsc}
\text{DSC}(X,Y) = 2\frac{|X \cap Y|}{|X|+|Y|} = \frac{2\text{TP}}{2\text{TP} + \text{FP} + \text{FN}} \, ,
\end{equation}
where TP, FP, FN stand for True Positives, False Positives, and False Negatives respectively.\\Since the goal was to maximize the Dice Score Coefficient, we defined the Dice Loss, to be minimized by the optimizer during training as shown in Equation~(\ref{eq:dcloss})
\begin{equation}
\label{eq:dcloss}
\text{Loss}(M_{gt} , M_{pred}) = 1 - \text{DSC}(M_{gt} , M_{pred})\, ,
\end{equation}
where $M_{gt}$, $M_{pred}$ is the ground truth target pupil mask and predicted pupil mask respectively.

\subsubsection{U-Net with Ellipse Regression (U-Net ER) Model}
\begin{figure} [!ht]
   \begin{center}
   \begin{tabular}{c} 
   \includegraphics[height=5cm]{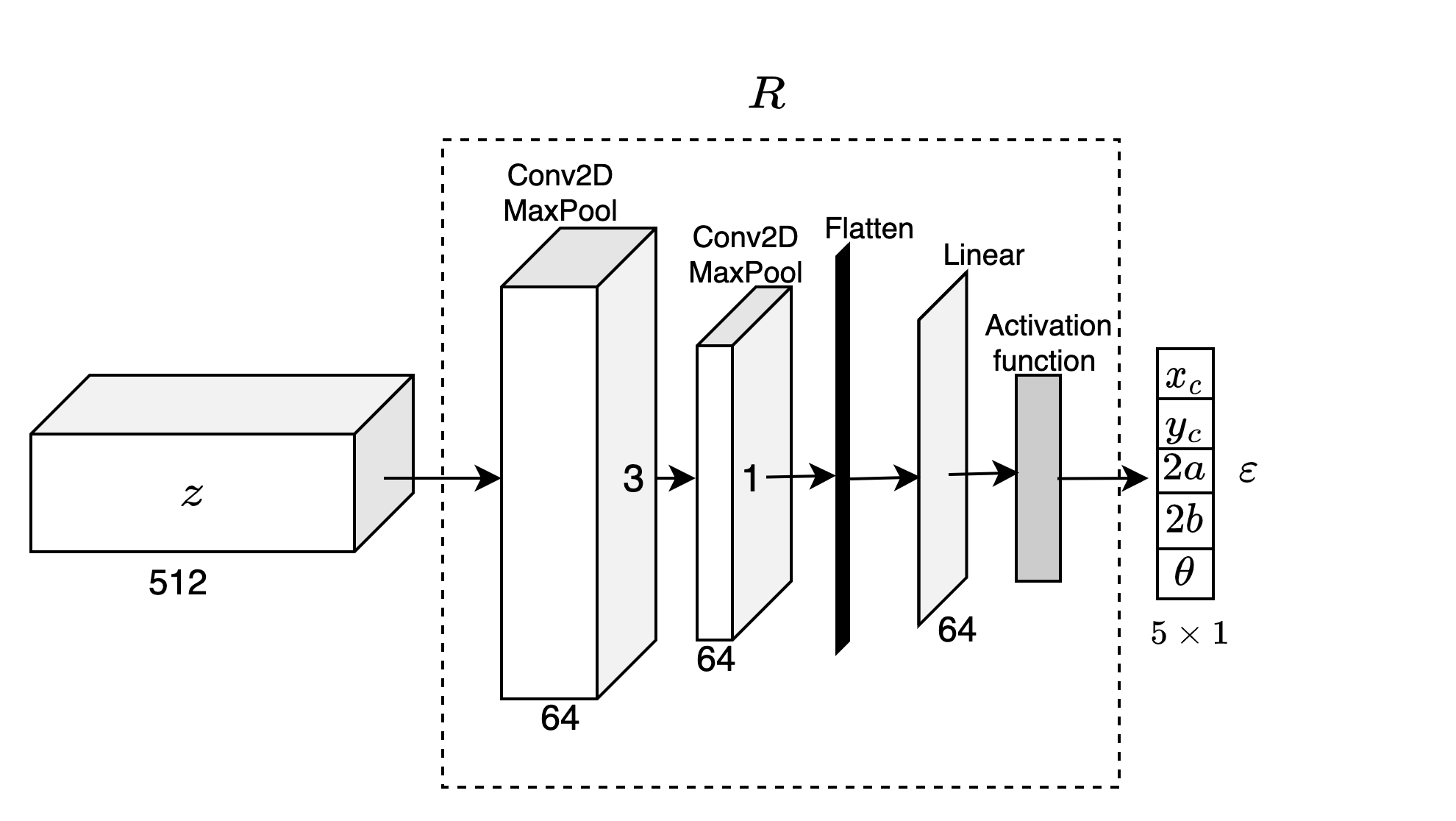}
   \end{tabular}
   \end{center}
   \caption[Regression Module]
   { \label{fig:regmod} Architecture used for evaluating ellipse regression ($R$) using bottleneck (latent) representation $z$.}
   \end{figure}

\begin{figure}[!h]
  \centering
  \begin{tabular}[b]{c}
    \includegraphics[width=.45\linewidth]{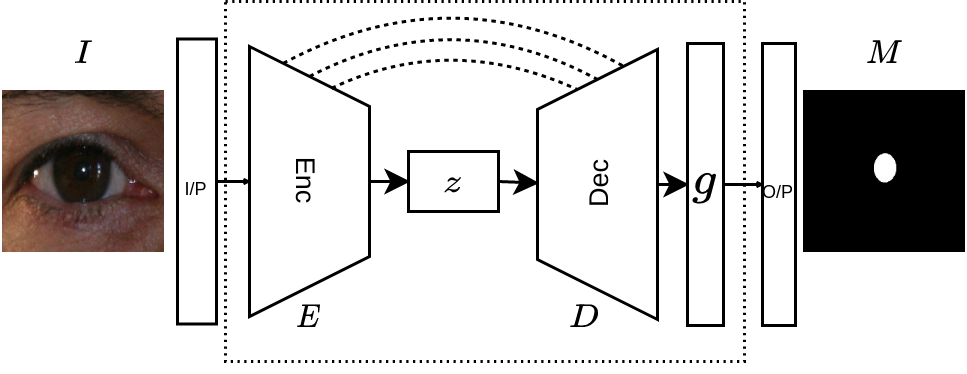} \\
    \small (a)
  \end{tabular} \enspace
  \\
  \begin{tabular}[b]{c}
    \includegraphics[width=.60\linewidth]{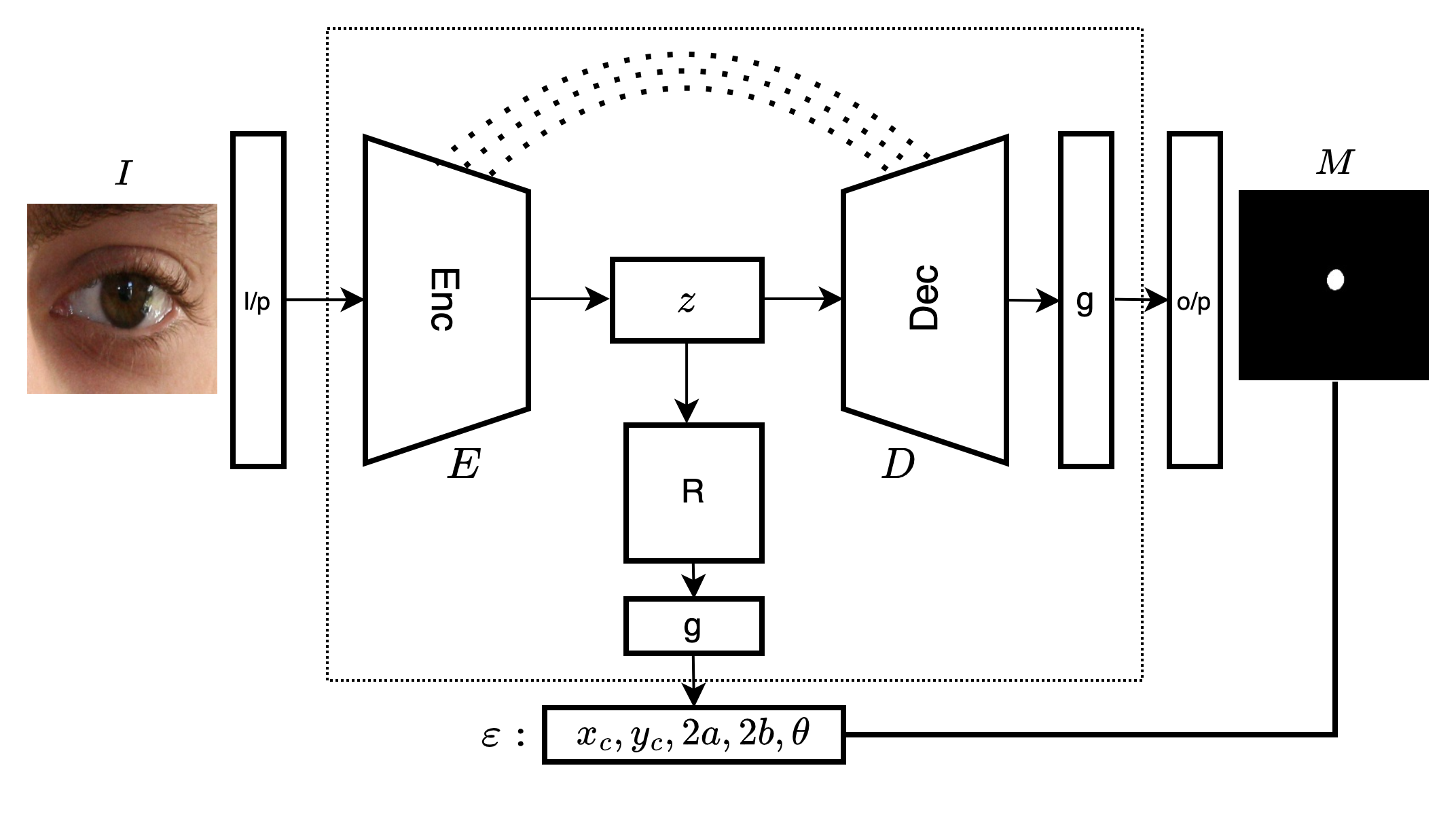} \\
    \small (b)
  \end{tabular}
 \caption[Schematic of the two neural network models] 
{ \label{fig:encdec} 
$I$ is the input RGB image. $M$ is the output predicted mask of the pupil. $E$ represents the encoder (downsampling), $D$ represents the decoder (upsampling), $z$ is the latent representation also referred to as bottleneck. Dotted lines from $E$ to $D$ are skip connections. $g$ is a non-linear activation function used to obtain $M$ and $\varepsilon$.
(a) Sample architecture for pupil segmentation only.
(b) Sample architecture for segmentation \& ellipse regression. $R$ is the ellipse regression module that can predict ellipse parameters $\varepsilon$ from the bottleneck represntation.}
\end{figure}

To obtain the ellipse parameters as well as the segmentation mask of pupil, we used the method described by Kothari et.al. \cite{kothari2021ellseg} to modify the network architecture as shown in Fig.~\ref{fig:encdec}(b).
The key idea here is to use a combined loss function for solving two learning tasks simultaneously: accurately identifying pixels representing the pupil ($M$) and regression parameters ($\varepsilon$) of the elliptical contour that fits the pixels identified as the pupil.
We introduce a regression module $R$ as shown in Fig.~\ref{fig:regmod} which is able to learn the ellipse parameters from the bottleneck representation $z$. For ellipse regression, we used the $L_{1}$ loss as it minimizes the absolute deviation between the predicted parameters and the ground truth. The joint loss function is defined in Equation~(\ref{eq:combloss})

\begin{equation}
\label{eq:combloss}
  \begin{array}{c}
\text{Loss}(M_{gt}, M_{pred}, \varepsilon_{gt}, \varepsilon_{pred}) = \text{DiceLoss}(M_{gt} , M_{pred}) + L_1(\varepsilon_{gt}, \varepsilon_{pred})\, \\
= 1 - \text{DSC}(M_{gt} , M_{pred}) + \sum |\varepsilon_{gt} - \varepsilon_{pred}|
    \end{array}
\end{equation}
where $\varepsilon_{gt}$, $\varepsilon_{pred}$ are the ground truth and predicted ellipse parameters respectively. 

\begin{figure*}[ht!]
\begin{minipage}[c]{0.95\linewidth}
\begin{center}
   \subfloat[Method 1]{%
      \includegraphics[ width=0.45\linewidth]{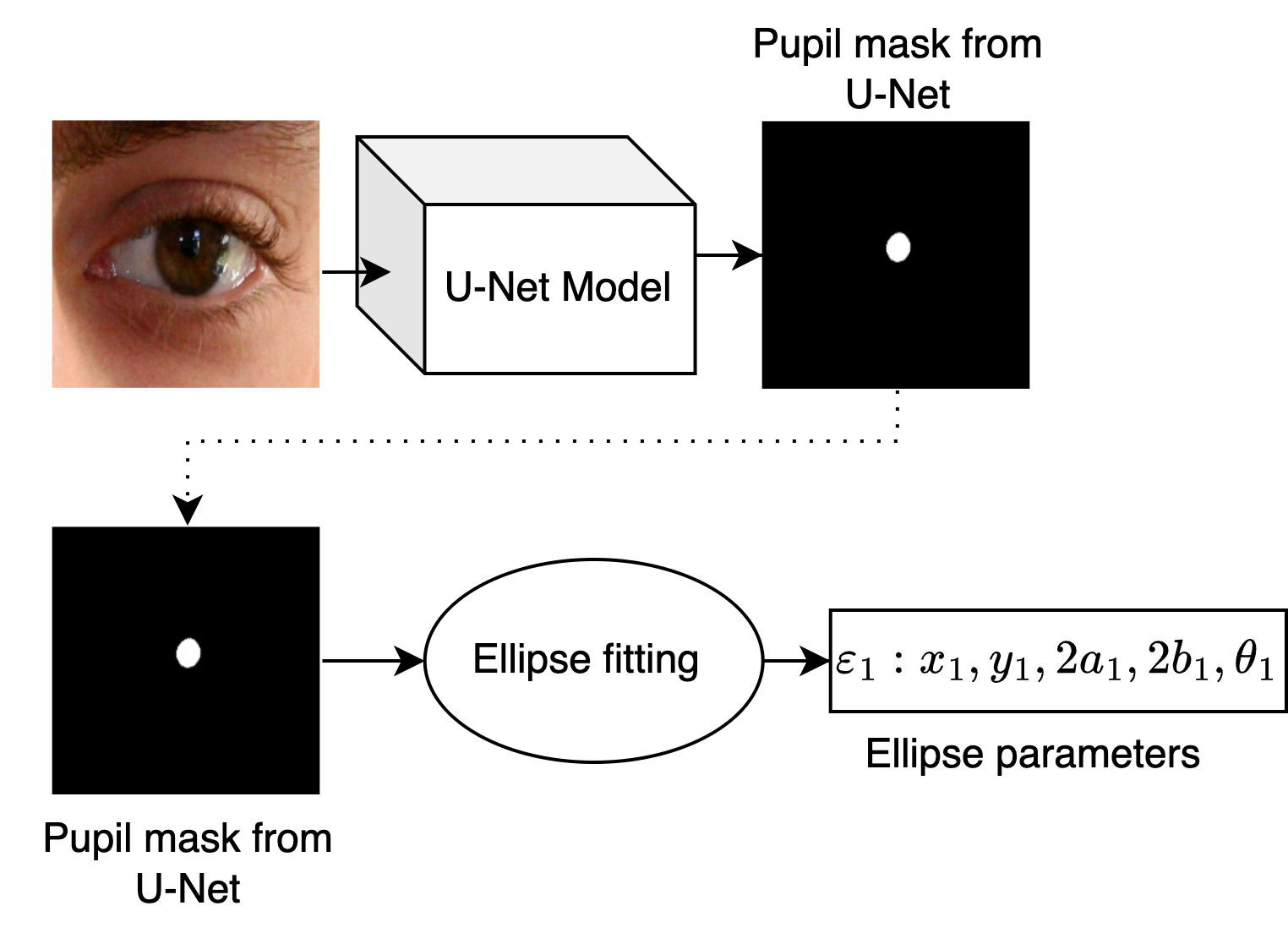}}
\hspace{\fill}
\end{center}
\end{minipage}
\newline
\begin{minipage}[c]{0.97\linewidth}
\begin{center}
   \subfloat[Method 2]{%
      \includegraphics[ width=0.46\linewidth]{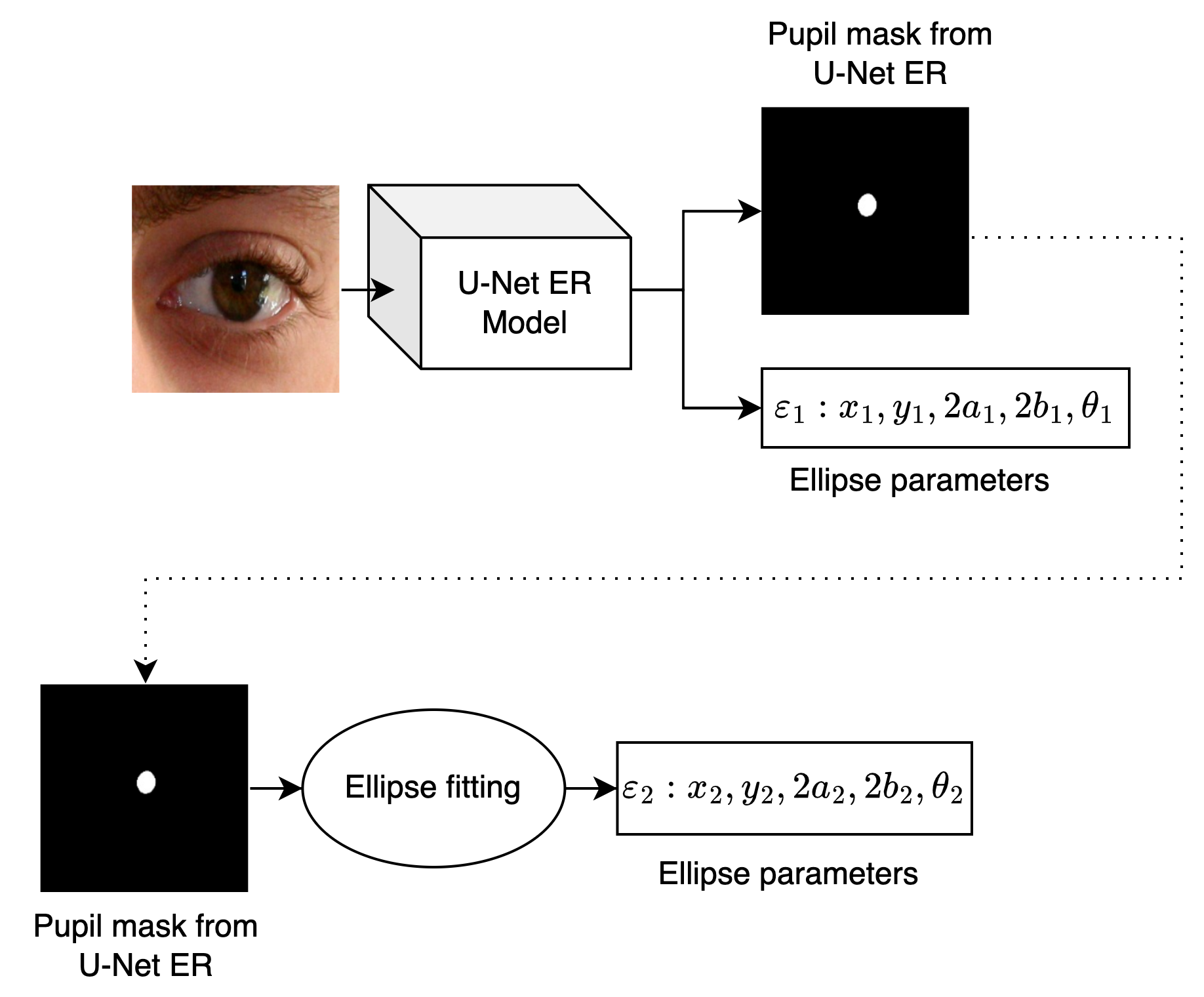}}
\hspace{\fill}
   \subfloat[Method 3]{%
      \includegraphics[ width=0.44\linewidth]{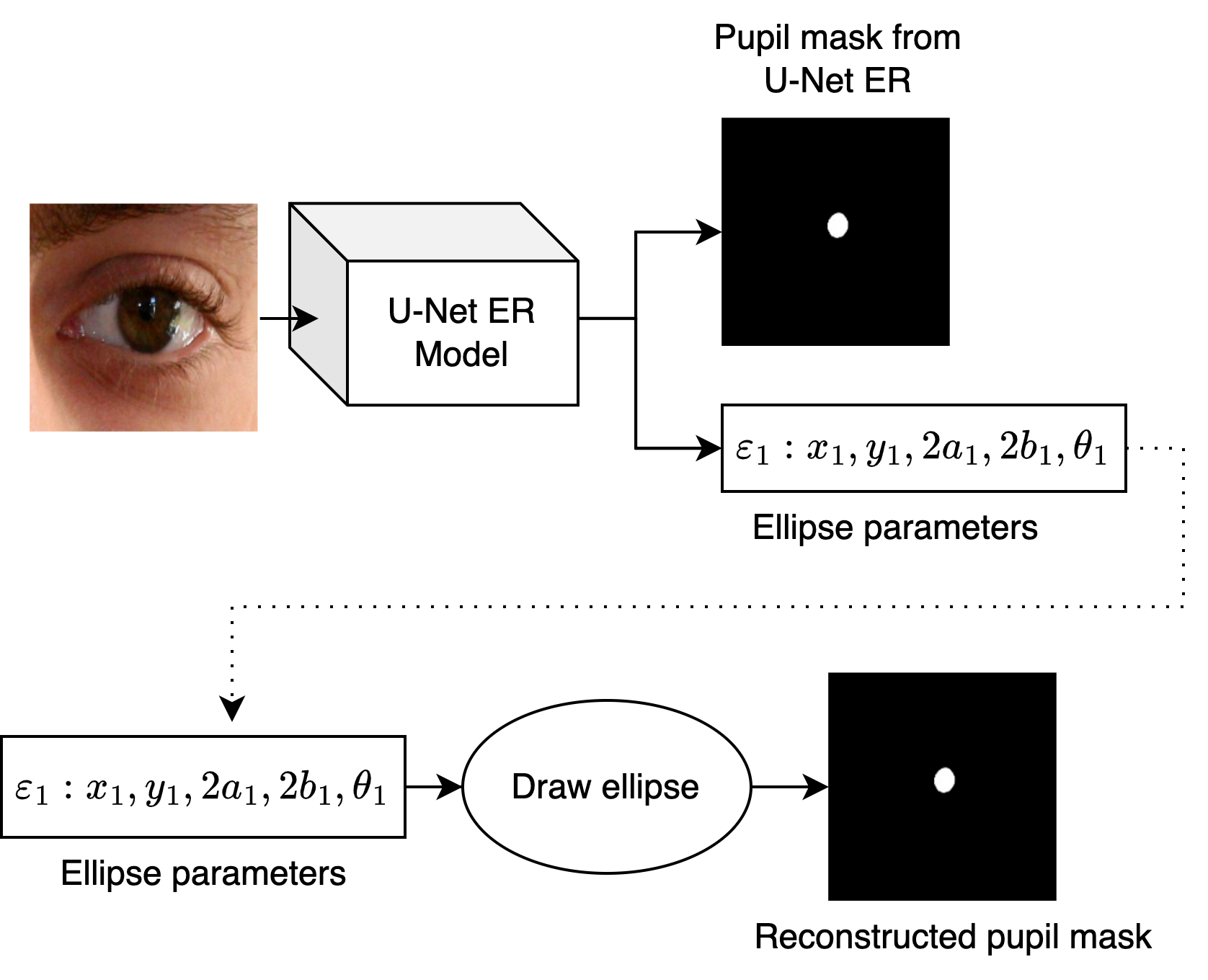}}\\
\end{center}
\end{minipage}
\caption{Visualization of the different methods of applying and evaluating U-Net and U-Net ER models. Pupil mask and parameters obtained after the dotted line are used for evaluation of each of the respective methods.}
    \label{fig:methods123}
\end{figure*}


\subsection{Training}
We used Pytorch\cite{paszke2019pytorch} to program the neural network architecture, training, and prediction pipeline in a GPU workstation with an NVIDIA A100 SXM graphics card. After applying augmentation techniques to the UBIRISv2 dataset, \redc{we obtained a total of 23431 input images and the corresponding ground truth pupil masks and ground truth pupil parameters}. Next, we split this dataset into training and validation sets in a 9:1 ratio. The input image originally of size (400, 300, 3) was resized to (224, 224, 3) and the pupil mask and ellipse parameters were adjusted accordingly.  To prepare the input data from the UBIRISv2 dataset, we performed normalization of both the input images and the ellipse parameters. Specifically, we scaled the pixel values of the images to a range of \redc{$[0, 1]$} by dividing each pixel value by 255. Additionally, the ellipse parameters were normalized by dividing the center coordinates $(x_c, y_c)$ and the major, minor axis $(2a, 2b)$ by 224, the dimension of the input images. This resulted in these parameters being scaled to the range of \redc{$[0, 1]$}. Finally, the orientation angle $\theta$ was divided by \redc{180$^\circ$}, which scaled it to the range of \redc{$[-1, 1]$}. A data generator was created to send mini-batches of data comprising of model specific inputs and outputs required by the model. Batch-size was set to 64. Models were trained for 100 epochs using the RMSprop optimizer\cite{tieleman2012lecture} with a learning rate of $1e^{-4}$, weight decay and momentum $= 1e^{-8},  0.9$ respectively. A cyclic learning rate scheduler was also used during training to adjust the learning rate of the optimizer. The CyclicLR scheduler from PyTorch was used with a minimum learning rate of $1e^{-4}$ and a maximum learning rate of $1e^{-3}$.  We used the Sigmoid function $g(u)=\frac{1}{1+e^{-u}}$ as the activation function $(g)$ after the decoder $(D)$ and ellipse regression module $(R)$ to obtain the pupil mask $M$ as well as ellipse parameters $\varepsilon$. After obtaining $\varepsilon$, we can rescale its original values by simple constant multiplication.

\subsection{Evaluation}

\redc{
To assess the performance of the proposed U-Net ER model in comparison to U-Net model, we conducted a comprehensive evaluation using three distinct methods, as illustrated in Fig. ~\ref{fig:methods123}. In Method 1, showcased in Fig. ~\ref{fig:methods123}(a), we employed a conventional ellipse fitting algorithm to extract ellipse parameters from the pupil mask generated by the U-Net model. The algorithm involved several steps, such as Gaussian blurring, thresholding, contour detection, and filtering based on aspect ratio and solidity criteria to identify the pupil contour. Subsequently, the \texttt{cv2.fitEllipse()} function from the OpenCV library\cite{opencv_library} was utilized to obtain the ellipse parameters. For Method 2, described in Fig. ~\ref{fig:methods123}(b), we obtained the predicted pupil mask directly from the U-Net ER model and applied the conventional ellipse fitting algorithm to derive the ellipse parameters $\varepsilon_2$ from the mask. In Method 3, as depicted in Fig. ~\ref{fig:methods123}(c), we directly used the ellipse parameters $\varepsilon_1$ obtained as output from the U-Net ER model. We used an ellipse drawing algorithm \texttt{cv2.ellipse()}\cite{opencv_library} on $\varepsilon_1$ to reconstruct the elliptical binary pupil mask for evaluation.
Comparing the results obtained from these three evaluation methods would enable gauging the efficacy and advantages of the U-Net ER model, which not only provides segmentation pupil masks but also directly outputs ellipse parameters.
To compare U-Net and U-Net ER, we measured the prediction times and evaluated the performance of the two models on UBIRISv2 test set and experimentally acquired images that were not used for training. We also evaluated the methods in terms of DSC and Accuracy metrics.} Accuracy is defined in Equation~(\ref{eq:acc})
\begin{equation}
\label{eq:acc}
\text{Accuracy}  = \frac{\text{TP} + \text{TN}}{\text{P} + \text{N}} \, ,
\end{equation}
where TP, TN, P, N stand for True Positives, True Negatives, Positives, and Negatives respectively based on the predicted and ground truth pupil masks.
As our future objective is to apply our approach to a video\redc{, frame by frame,} and plot the pupillogram \cite{wilhelm2003clinical}, \redc{it was important to measure the errors in estimating pupil diameter}. This is important as the pupillogram represents changes in pupil diameter over time, and accurate estimation of pupil diameter is necessary for meaningful clinical analysis. Therefore, we aimed to ensure that our approach provides reliable estimates of pupil diameter, which can be used to generate accurate pupillograms from video data. \redc{Pupil radius is defined as the }average length of the semi-axes $(\frac{a + b}{2})$, and the pupil diameter is $\text{PD}=(a + b)$ and the pupil diameter error is calculated as $|\text{PD}_{gt}-\text{PD}_{pred}|$, where $\text{PD}_{gt}$ and $\text{PD}_{pred}$ denote the ground truth and predicted pupil diameters respectively. \redc{Pupil position is defined by the coordinates of the center $(x_c, y_c)$ of the fitted ellipse. The deviation of these coordinates from ground truth data is referred to as Pupil Position Error and calculated with respect to $x$ and $y$ axes as $|x_{c\_gt} - x_{c\_pred}|$ and $|y_{c\_gt} - y_{c\_pred}|$ respectively where $(x_{c\_gt}, y_{c\_gt})$ and $(x_{c\_pred}, y_{c\_pred})$ denote the ground truth and predicted pupil positions respectively.}

\subsection{Experimentally acquired images}
\redc{To mimic the field conditions and evaluate the robustness of model predictions on out of distribution test samples}, pupillary images from 5 self-reported healthy volunteers were collected. Participants were seated facing a slit-lamp and a diffuse illumination setting was used to illuminate the eye. Images were sequentially captured with a commercially available Android smartphone (Xiaomi Redmi Note 10, Model no: M2101K7AI) using the inbuilt camera application. The camera was positioned at a distance of 15-18 cm from the eye, and adjusted so that the eye is roughly centered in the image. Ground truth pupil masks were created manually  by a clinically experienced author (JJB) using ImageJ software \cite{schneider2012nih} (version 1.51w) and ellipse parameters were extracted using the least squares algorithm.

\section{RESULTS AND DISCUSSION}
Table~\ref{tab:modelarch} shows a comparison of the two models in terms of computational performance. We saw that obviating the additional ellipse fitting step is useful as it reduces the total computation time. It is also important to note that since these models will eventually be deployed on a smartphone device, saving time and computational resource is critical.
\begin{table}[ht!]
\caption{Comparison in terms of computational performance} 
\label{tab:modelarch}
\begin{center}       
\begin{tabular}{|l|c|c|}
\hline
\rule[-1ex]{0pt}{3.5ex} \redc{\textbf{Criteria (Units)}}  & \redc{\textbf{U-Net}} & \redc{\textbf{U-Net ER}}
\\
\hline
\rule[-1ex]{0pt}{3.5ex}  Number of trainable model parameters (approx in million) & 7.8 &
8.1\\
\hline
\rule[-1ex]{0pt}{3.5ex}  Model memory (MB) & 30 & 31  \\
\hline
\rule[-1ex]{0pt}{3.5ex}  Time for neural network prediction (ms/frame) & 15 & 17   \\
\hline 
\rule[-1ex]{0pt}{3.5ex}  Time for ellipse fitting (ms/frame) & 8 & 0 \\
\hline 
\rule[-1ex]{0pt}{3.5ex}  Total Time (ms) & 23 & 17   \\
\hline 
\end{tabular}
\end{center}
\end{table}
\begin{table}[!ht]
\centering
\caption{Evaluation of UBIRISv2 ($n = 30 $)}
\label{tab:modelubacc}
\begin{tabular}{|c|c|c|c|}
\hline
\textbf{Parameters (Units)} & \multicolumn{3}{c|}{\textbf{UBIRISv2}} \\ \hline
\textbf{$( \mu , \sigma )$} & \textbf{Method 1} & \textbf{Method 2} & \textbf{Method 3} \\ \hline
$\text{DSC}$ & ( 0.987 , 0.007 ) & ( 0.992 , 0.006 ) & ( 0.840 , 0.065 ) \\ \hline
$\text{Accuracy}$ & ( 0.998 , 0.001 ) & ( 0.998 , 0.001 ) & ( 0.998 , 0.001 ) \\ \hline
$\text{PD}$ error (pixels) & ( 0.071 , 0.053 ) & ( 0.010 , 0.072 ) & ( 1.956 , 0.353 ) \\ \hline
$x_{c}$ error (pixels) & ( 0.047 , 0.038 ) & ( 0.039 , 0.026 ) & ( 0.277 , 0.189 ) \\ \hline
$y_{c}$ error (pixels) & ( 0.061 , 0.043 ) & ( 0.075 , 0.065 ) & ( 0.434 , 0.372 ) \\ \hline
\end{tabular}
\end{table}

\begin{table}[!ht]
\centering
\caption{Evaluation of Experimentally Acquired Images ($n = 30$)    }
\label{tab:modelubaccexpt}
\begin{tabular}{|c|c|c|c|}
\hline
\textbf{Parameters (Units)} & \multicolumn{3}{c|}{\textbf{Experimentally Acquired Images}} \\ \hline
\textbf{$( \mu, \sigma )$} & \textbf{Method 1} & \textbf{Method 2} & \textbf{Method 3} \\ \hline
$\text{DSC}$ & ( 0.681 , 0.280 ) & ( 0.780 , 0.147 ) & ( 0.653 , 0.123 ) \\ \hline
$\text{Accuracy}$ & ( 0.998 , 0.001 ) & ( 0.996 , 0.001 ) & ( 0.997 , 0.001 ) \\ \hline
$\text{PD}$ error (pixels) & ( 3.110 , 2.318 ) & ( 2.335 , 0.835 ) & ( 1.226 , 1.126 ) \\ \hline
$x_{c}$ error (pixels) & ( 8.815 , 31.554 ) & ( 0.332 , 0.510 ) & ( 2.204 , 1.436 )\\ \hline
$y_{c}$ error (pixels) & ( 7.728 , 25.199 ) & ( 0.854 , 2.194 ) & ( 2.147 , 1.189 ) \\ \hline
\end{tabular}
\end{table}

\begin{figure}[!ht]
  \centering
  \begin{tabular}[b]{c}
    \includegraphics[width=.42\linewidth]{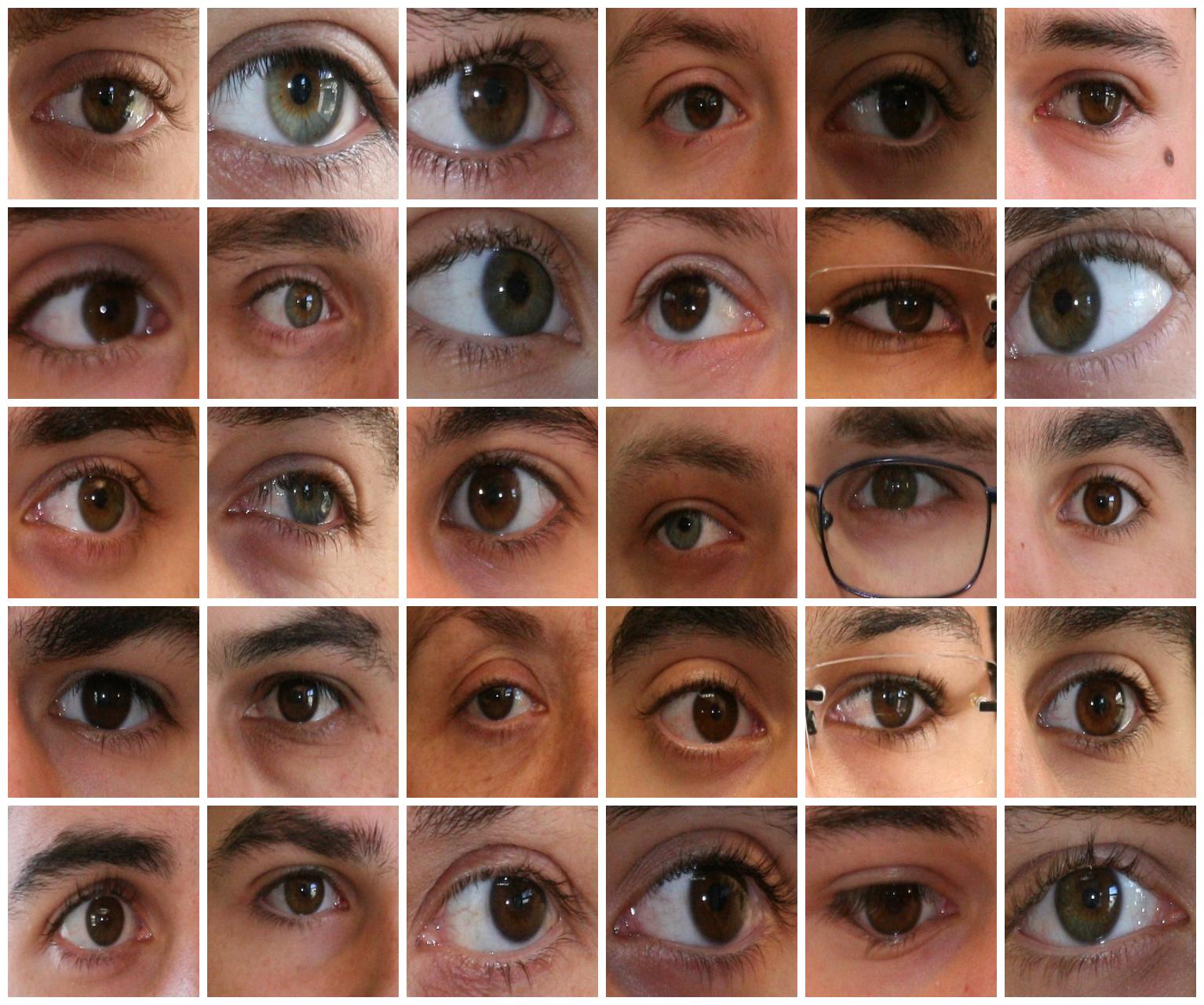} \\
    \small (a)
  \end{tabular} \enspace
  \begin{tabular}[b]{c}
    \includegraphics[width=.42\linewidth]{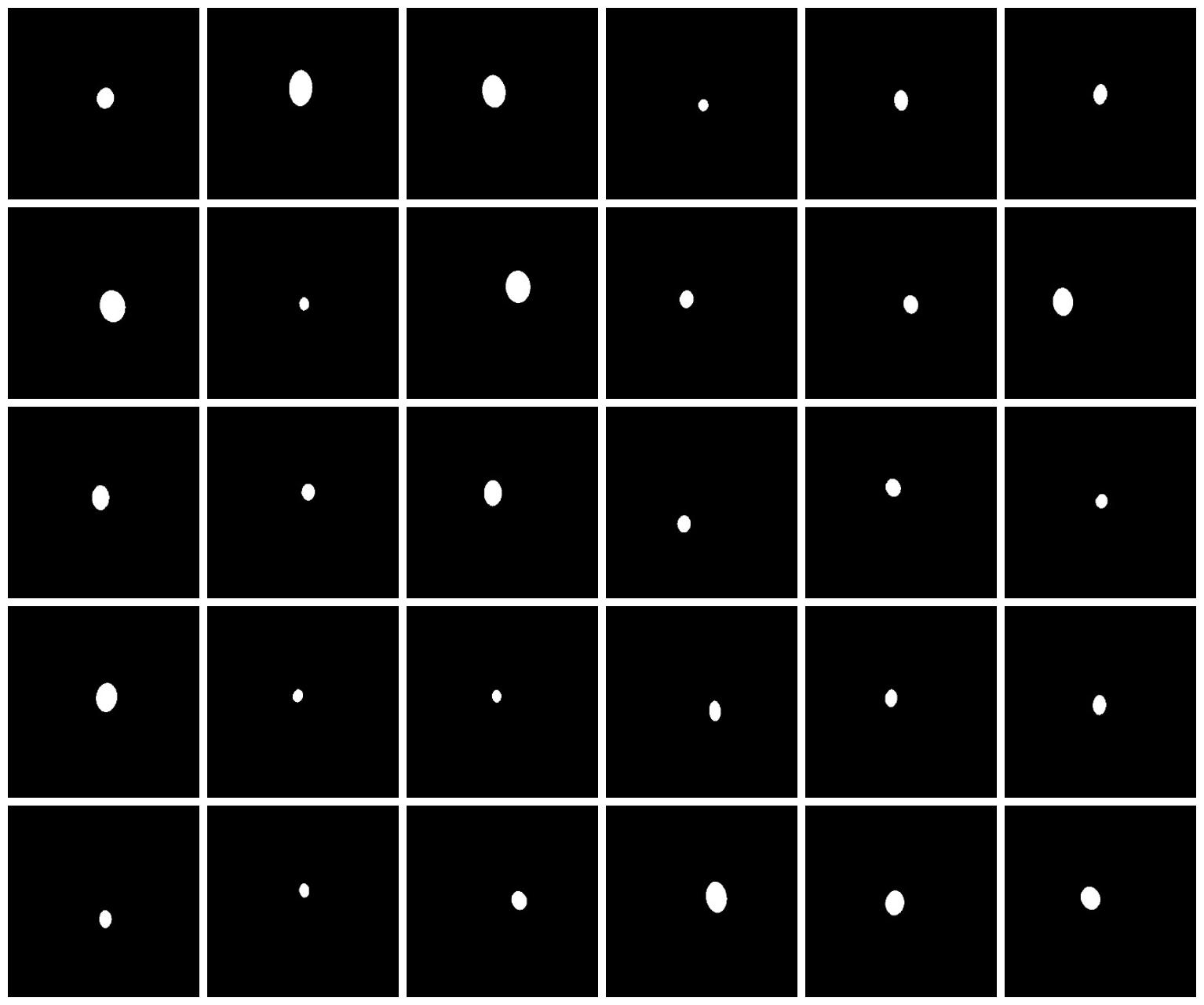} \\
    \small (b)
  \end{tabular} \enspace
  \\
   \begin{tabular}[b]{c}
    \includegraphics[width=.42\linewidth]{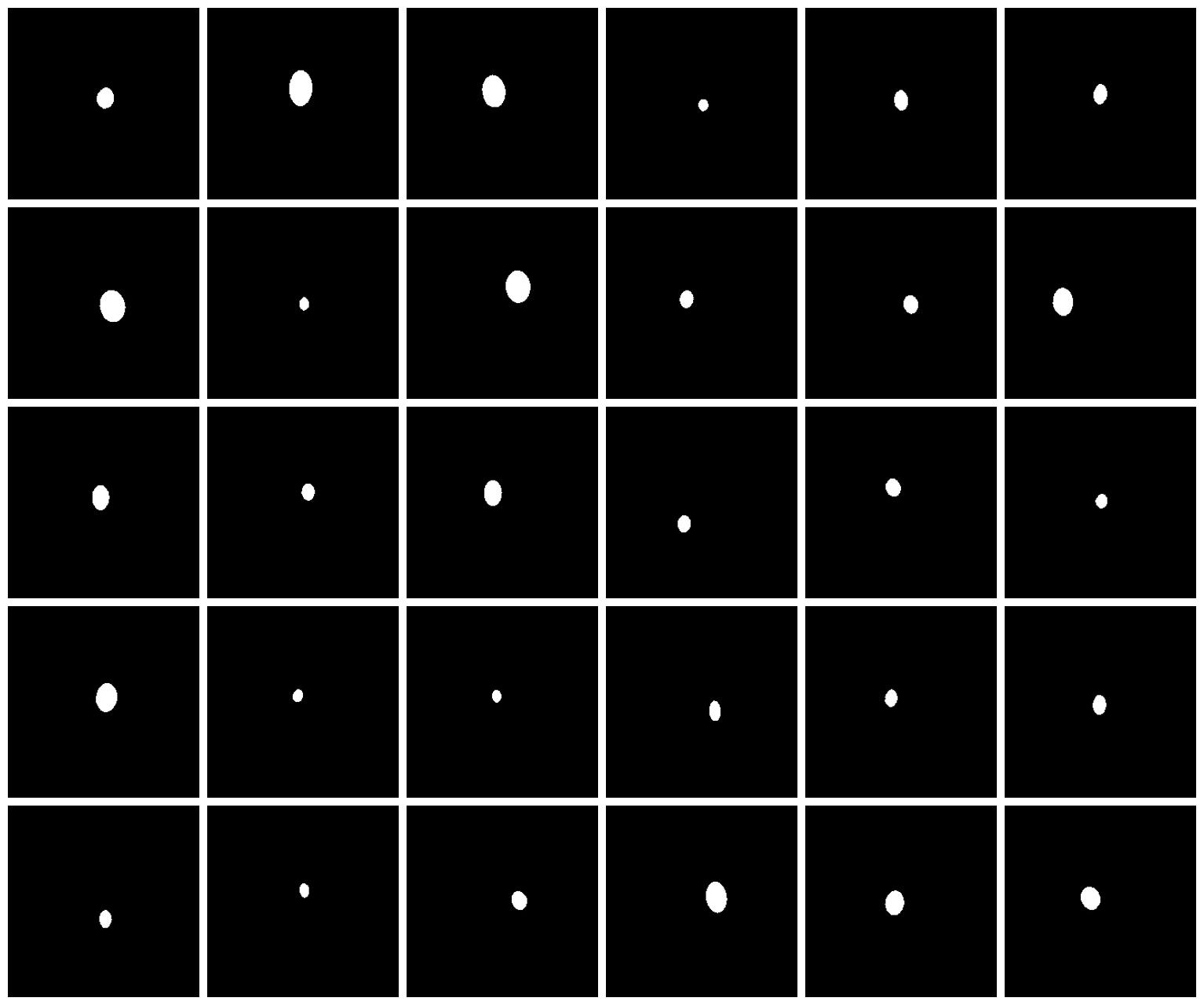} \\
    \small (c)
  \end{tabular} \enspace
  \begin{tabular}[b]{c}
    \includegraphics[width=.42\linewidth]{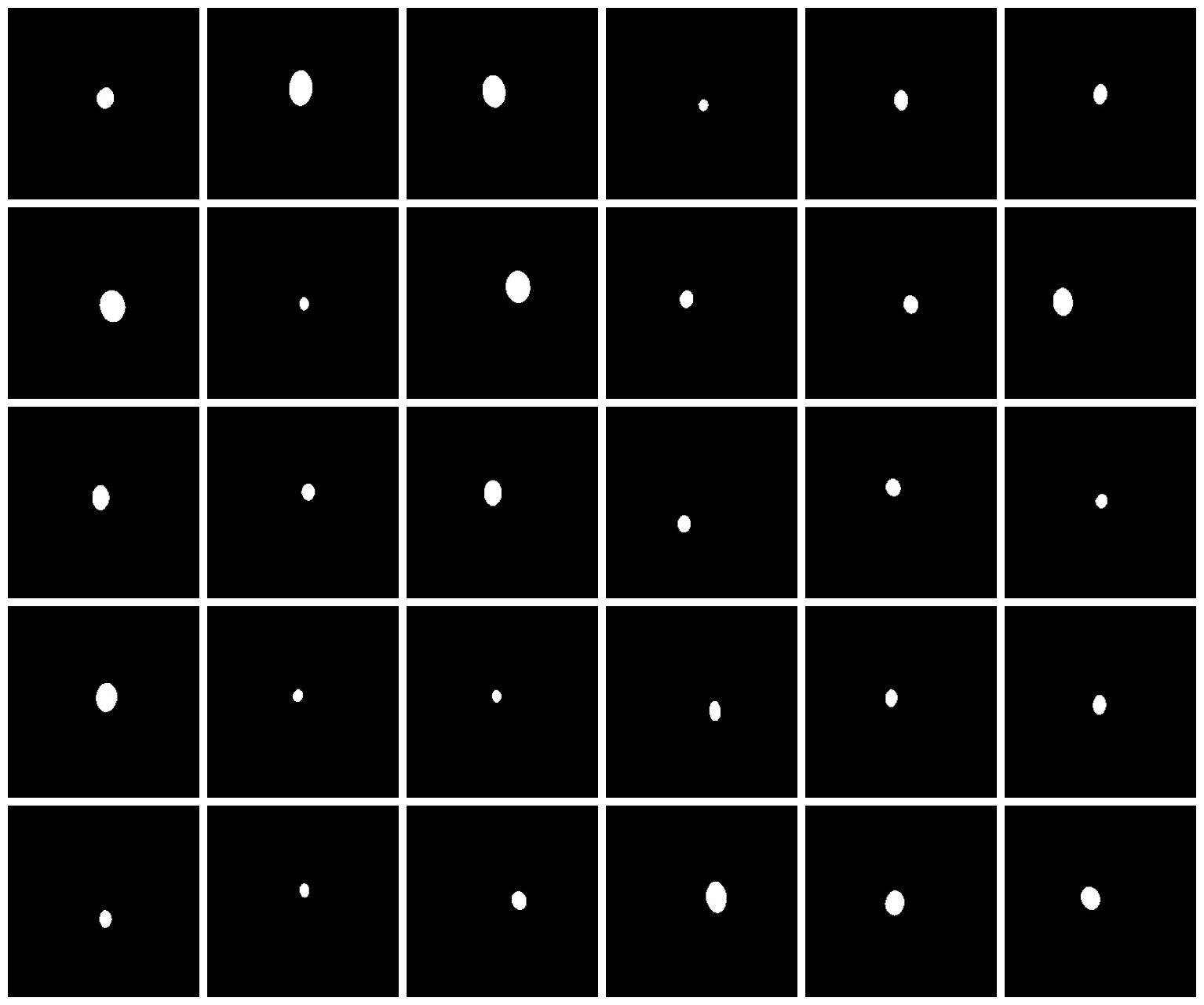} \\
    \small (d)
  \end{tabular}
  \caption{\label{fig:eye1} Results on UBIRISv2. (a) RGB images from UBIRISv2, (b) Ground Truth Pupil Masks, (c) Predictions from U-Net, (d)  Predictions from U-Net ER.}
\end{figure}
\begin{figure}[!ht]
  \centering
  \begin{tabular}[b]{c}
    \includegraphics[width=.42\linewidth]{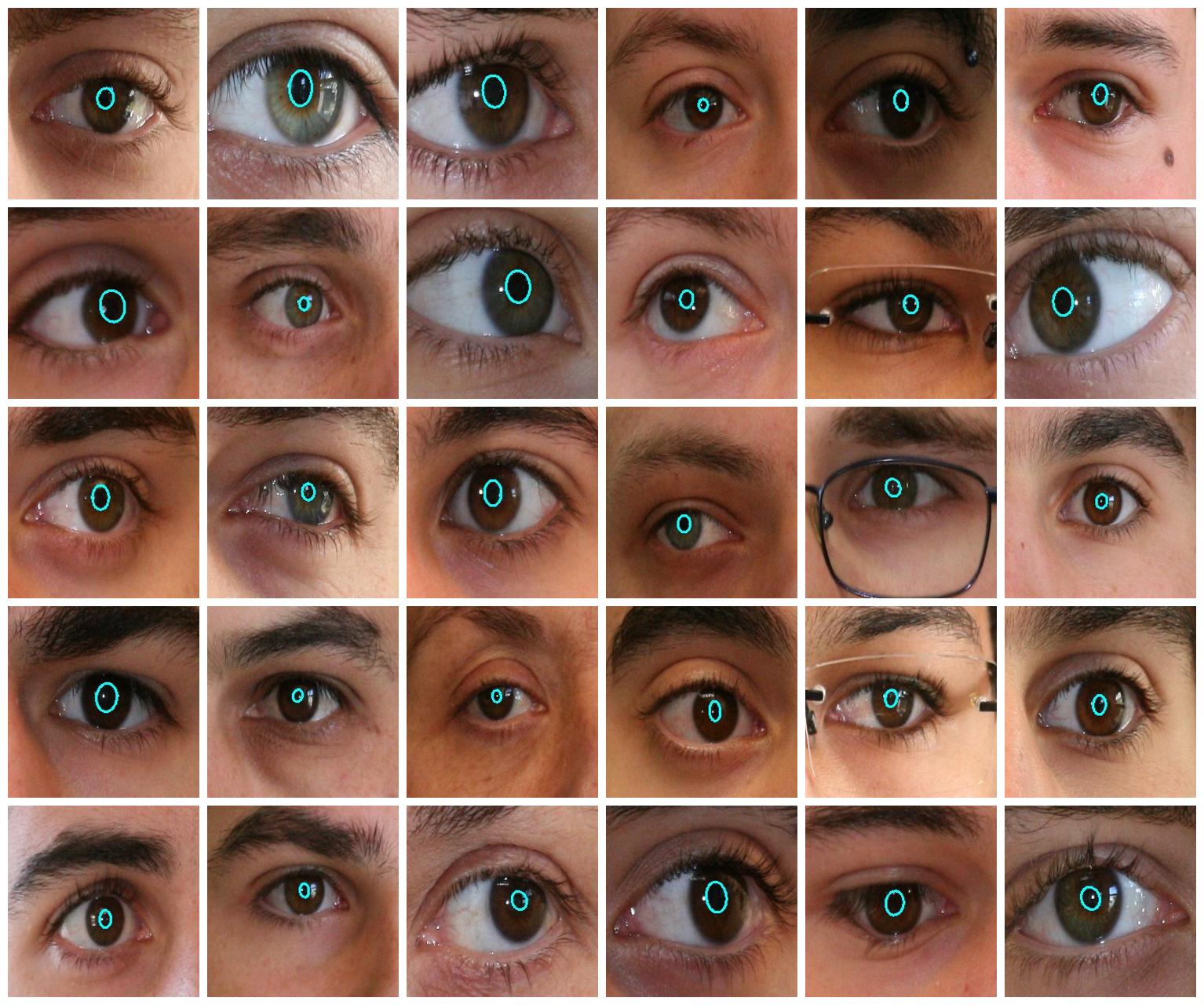} \\
    \small (a)
  \end{tabular} \enspace
  \begin{tabular}[b]{c}
    \includegraphics[width=.42\linewidth]{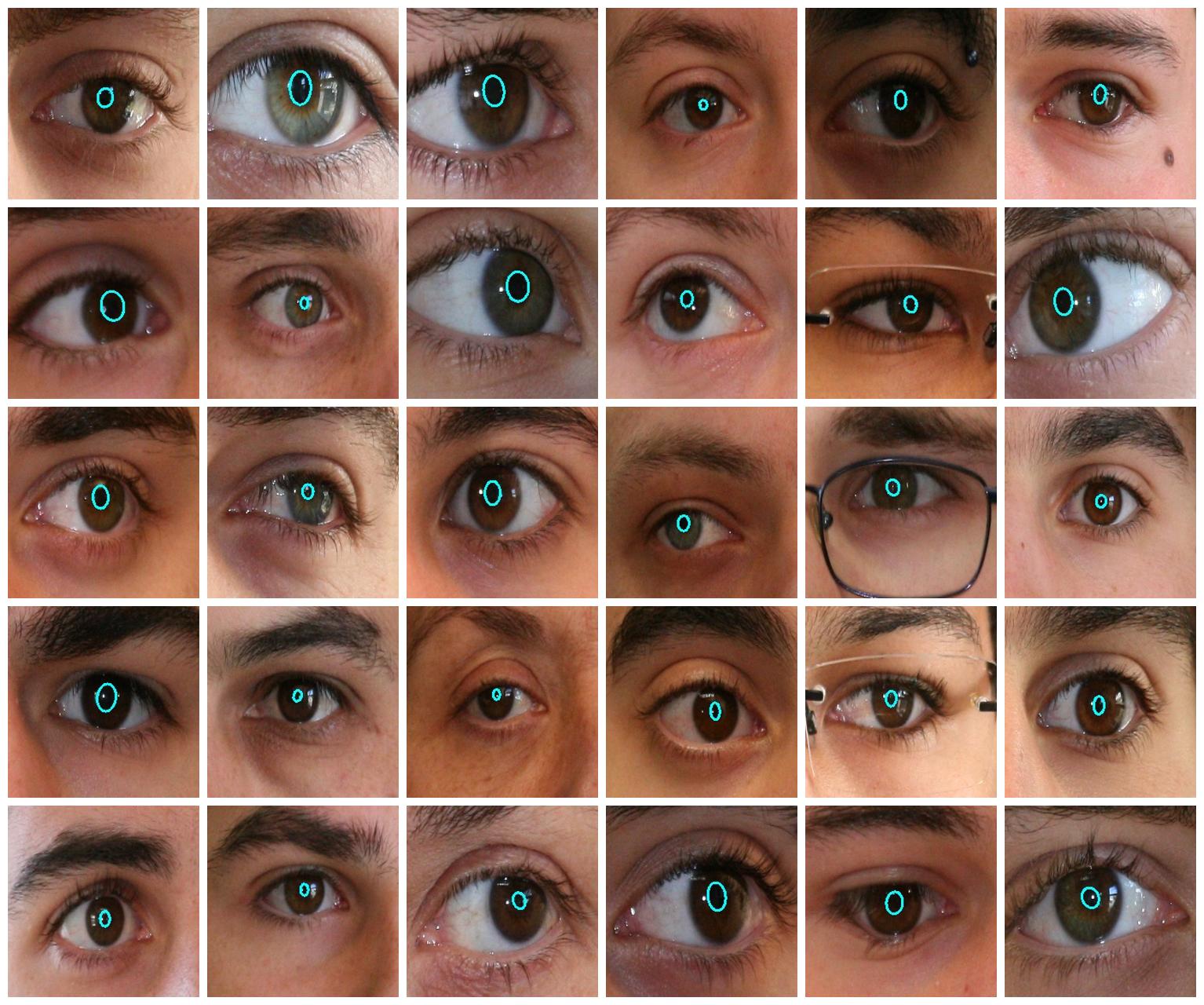} \\
    \small (b)
  \end{tabular}
  \caption{\label{fig:eye2} Results on UBIRISv2. (a) \redc{Fitted ellipse boundary using $\varepsilon_{1}$ from U-Net (Method 1)}, (b) \redc{Fitted ellipse boundary using $\varepsilon_{2}$ from U-Net ER (Method 2)}.}
\end{figure}

\redc{Tables ~\ref{tab:modelubacc} and ~\ref{tab:modelubaccexpt} presents the summary statistics ( $\mu, \sigma$ stand for mean and standard deviation) for the U-Net and U-Net ER models, which were evaluated on both the UBIRISV2 and experimentally acquired images datasets respectively}. To get an estimate of the performance of the trained models on unseen data, we sampled 30 random images $(n)$ from the UBIRISv2 dataset and experimentally acquired data each and ran predictions. Figs.~\ref{fig:eye1},~\ref{fig:eye2} illustrates the performance visually. We saw that the U-Net ER performs better than the U-Net. 

\begin{figure}[!ht]
  \centering
  \begin{tabular}[b]{c}
    \includegraphics[width=.42\linewidth]{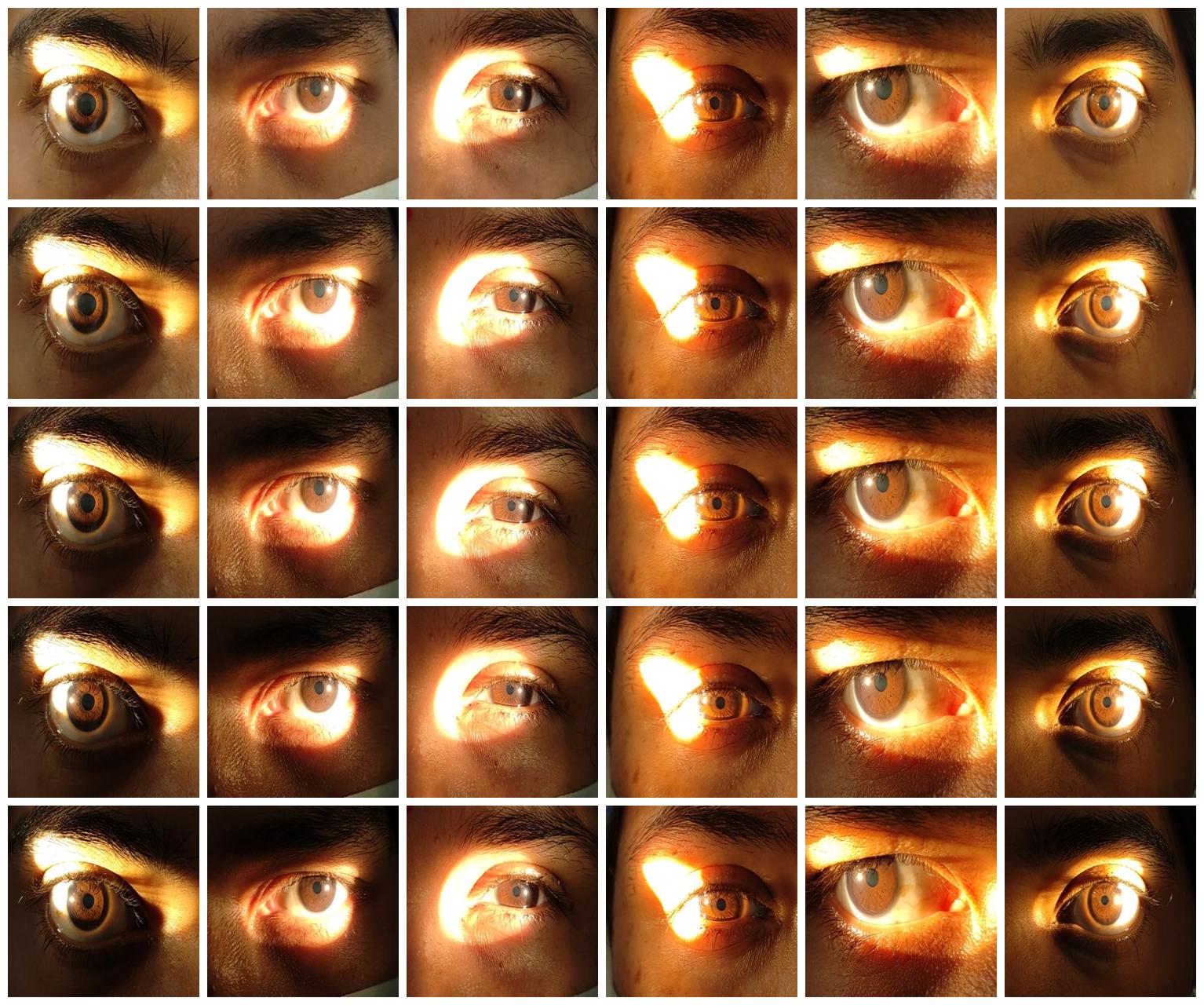} \\
    \small (a)
  \end{tabular} \enspace
  \begin{tabular}[b]{c}
    \includegraphics[width=.42\linewidth]{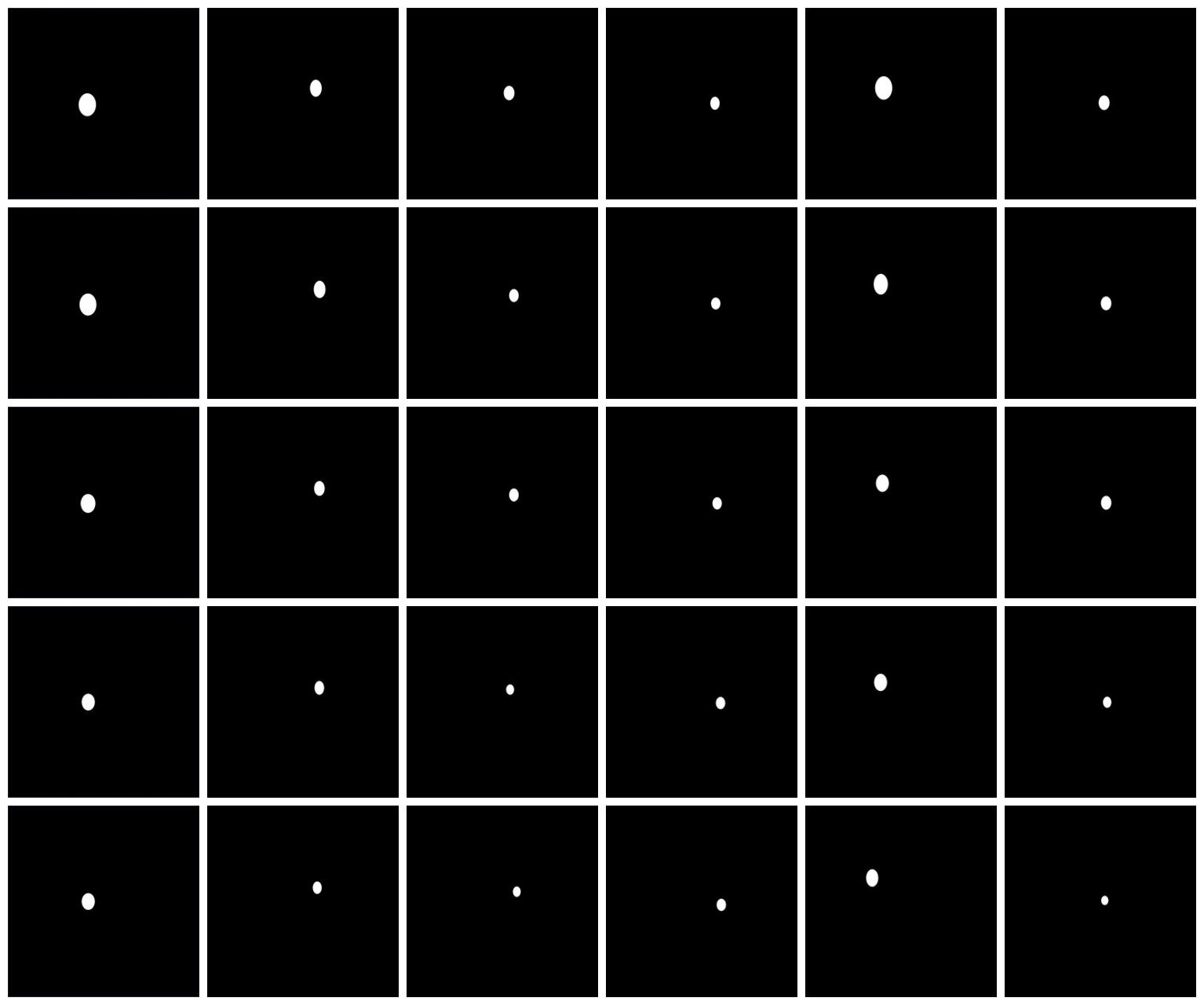} \\
    \small (b)
  \end{tabular} \enspace
  \\
   \begin{tabular}[b]{c}
    \includegraphics[width=.42\linewidth]{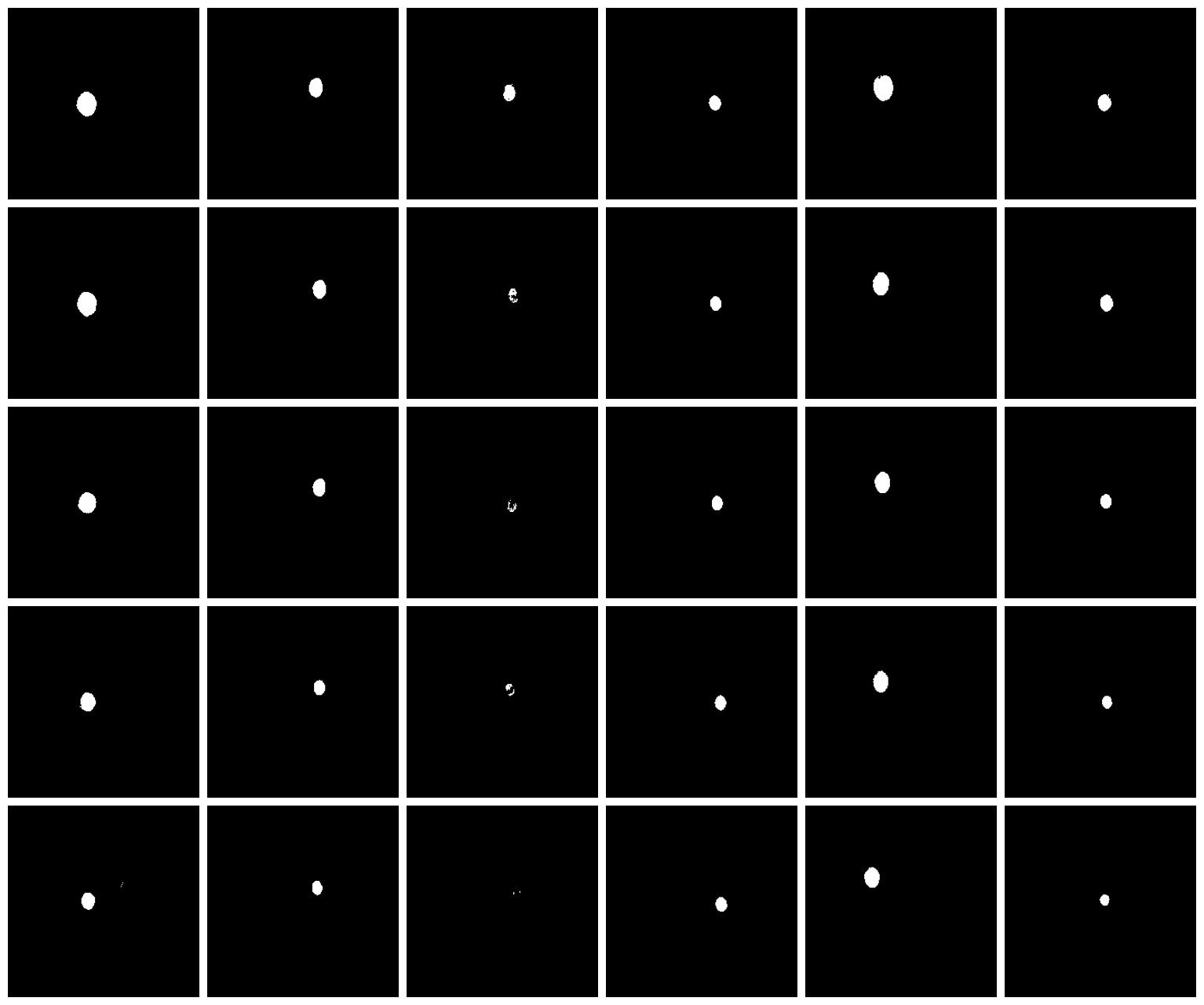} \\
    \small (c)
  \end{tabular} \enspace
   \begin{tabular}[b]{c}
    \includegraphics[width=.42\linewidth]{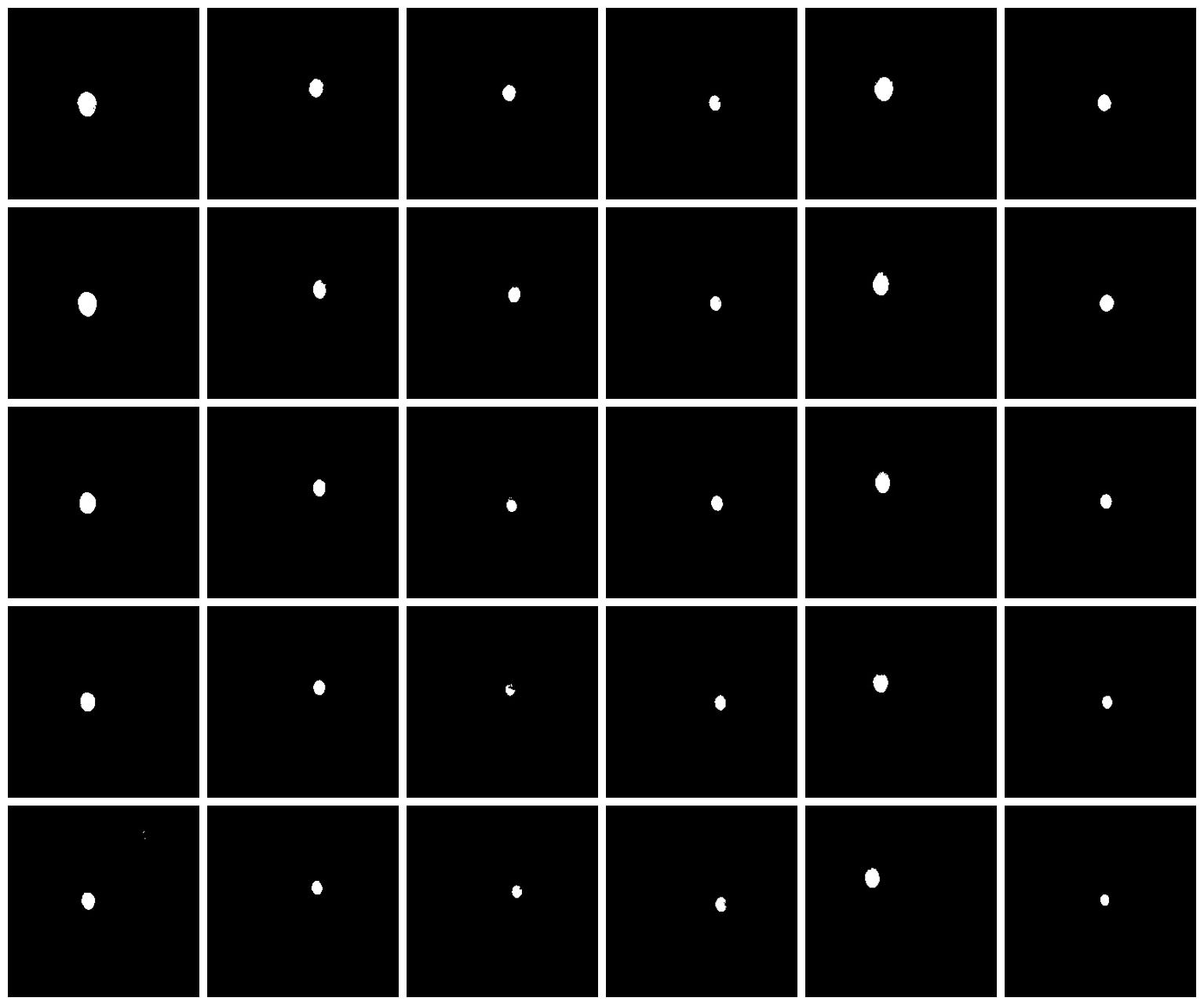} \\
    \small (d)
  \end{tabular}
  \caption{\label{fig:eye3} Results on Experimentally Acquired Images (a) RGB Images, (b) Ground Truth Pupil Masks, (c) Predictions from U-Net, (d) Predictions from U-Net ER.}
\end{figure}

\begin{figure}[!ht]
  \centering
  \begin{tabular}[b]{c}
    \includegraphics[width=.42\linewidth]{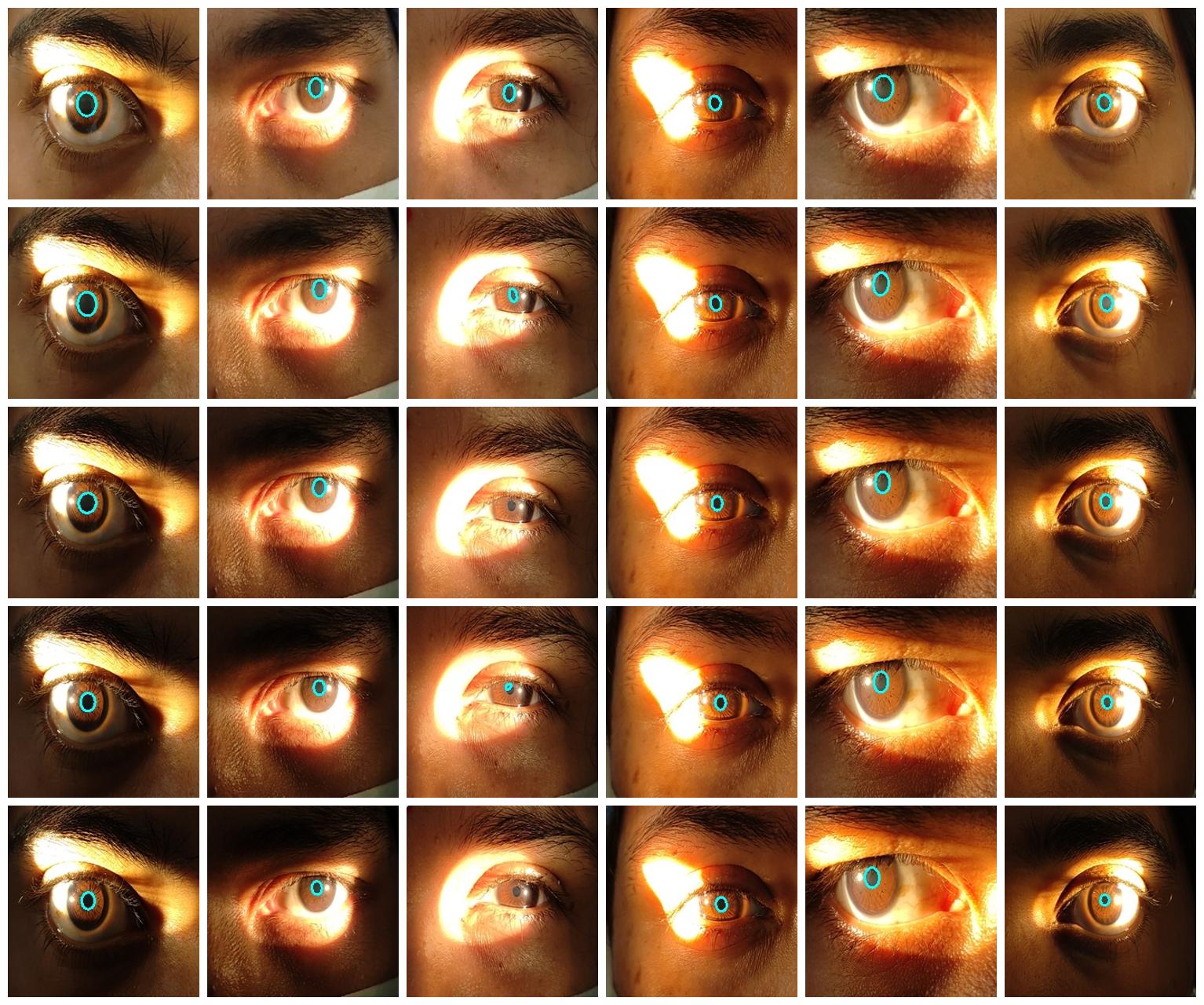} \\
    \small (a)
  \end{tabular} \enspace
  \begin{tabular}[b]{c}
    \includegraphics[width=.42\linewidth]{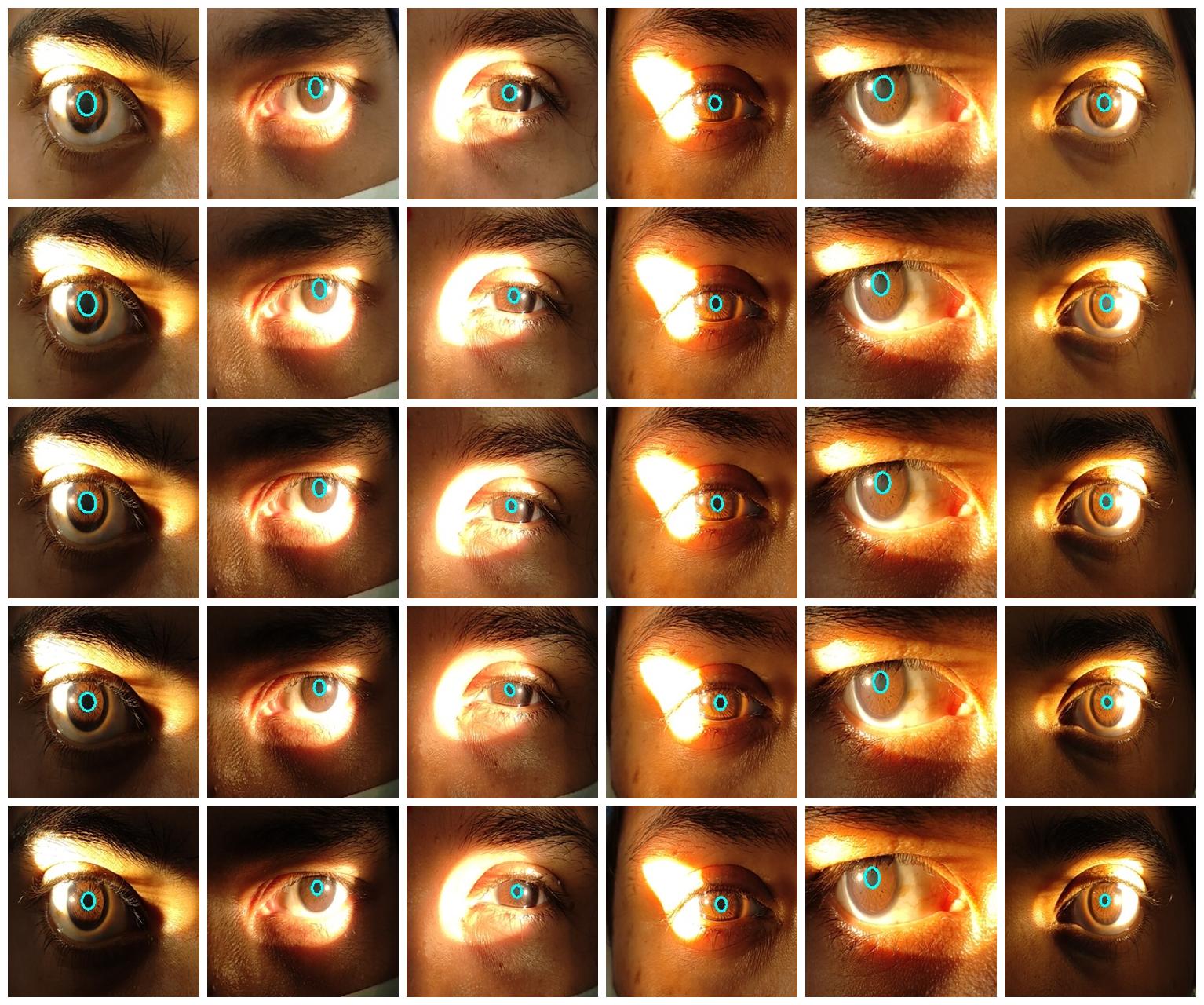} \\
    \small (b)
  \end{tabular}
  \caption{\label{fig:eye4}  Results on Experimentally Acquired Images. (a) \redc{Fitted ellipse boundary using $\varepsilon_{1}$ from U-Net (Method 1)}, (b) \redc{Fitted ellipse boundary using $\varepsilon_{2}$ from U-Net ER (Method 2)}.}
\end{figure}

\begin{figure}[!ht]
  \centering
  \begin{tabular}[b]{c}
    \includegraphics[width=.42\linewidth]{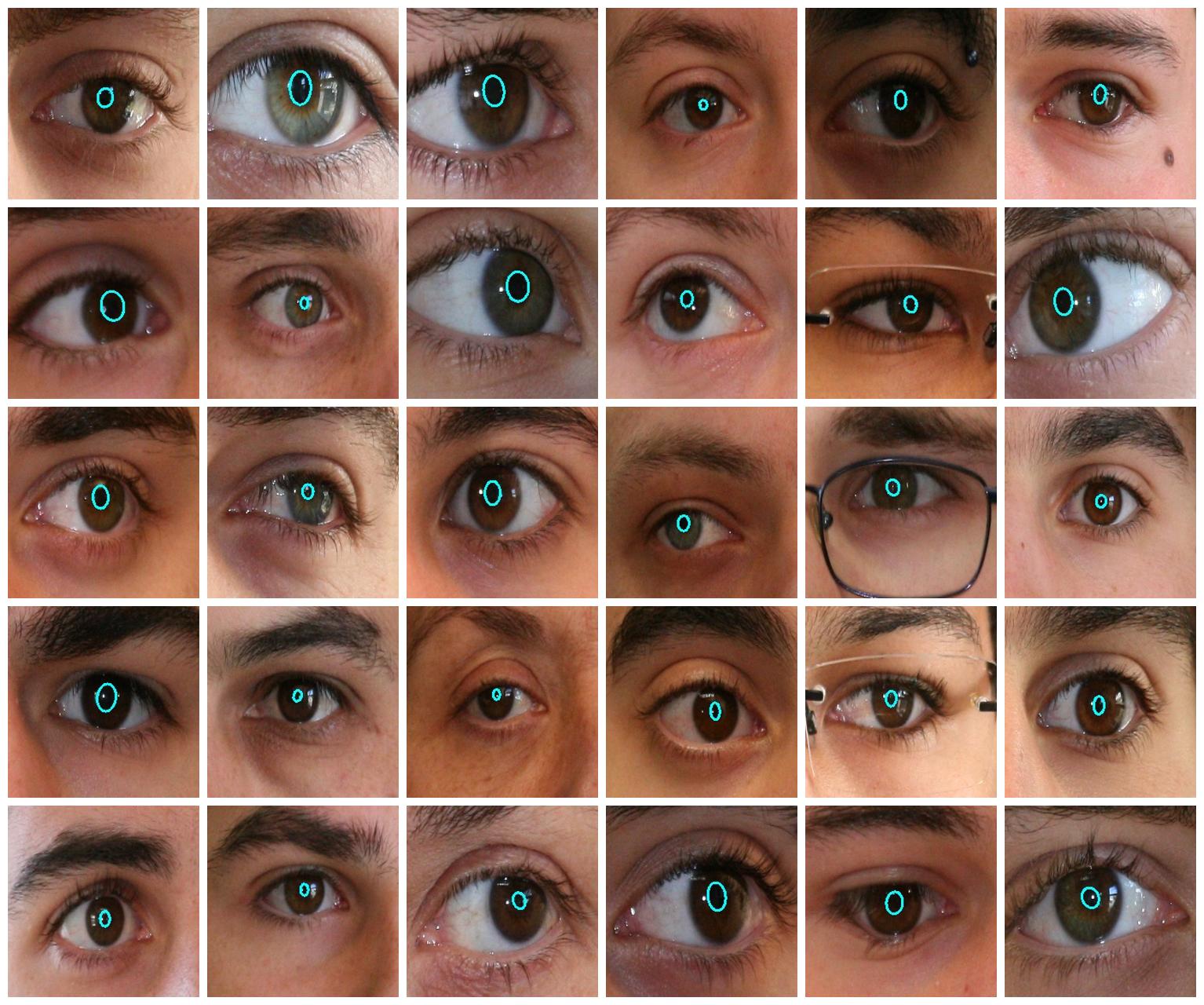} \\
    \small (a)
  \end{tabular} \enspace
  \begin{tabular}[b]{c}
    \includegraphics[width=.42\linewidth]{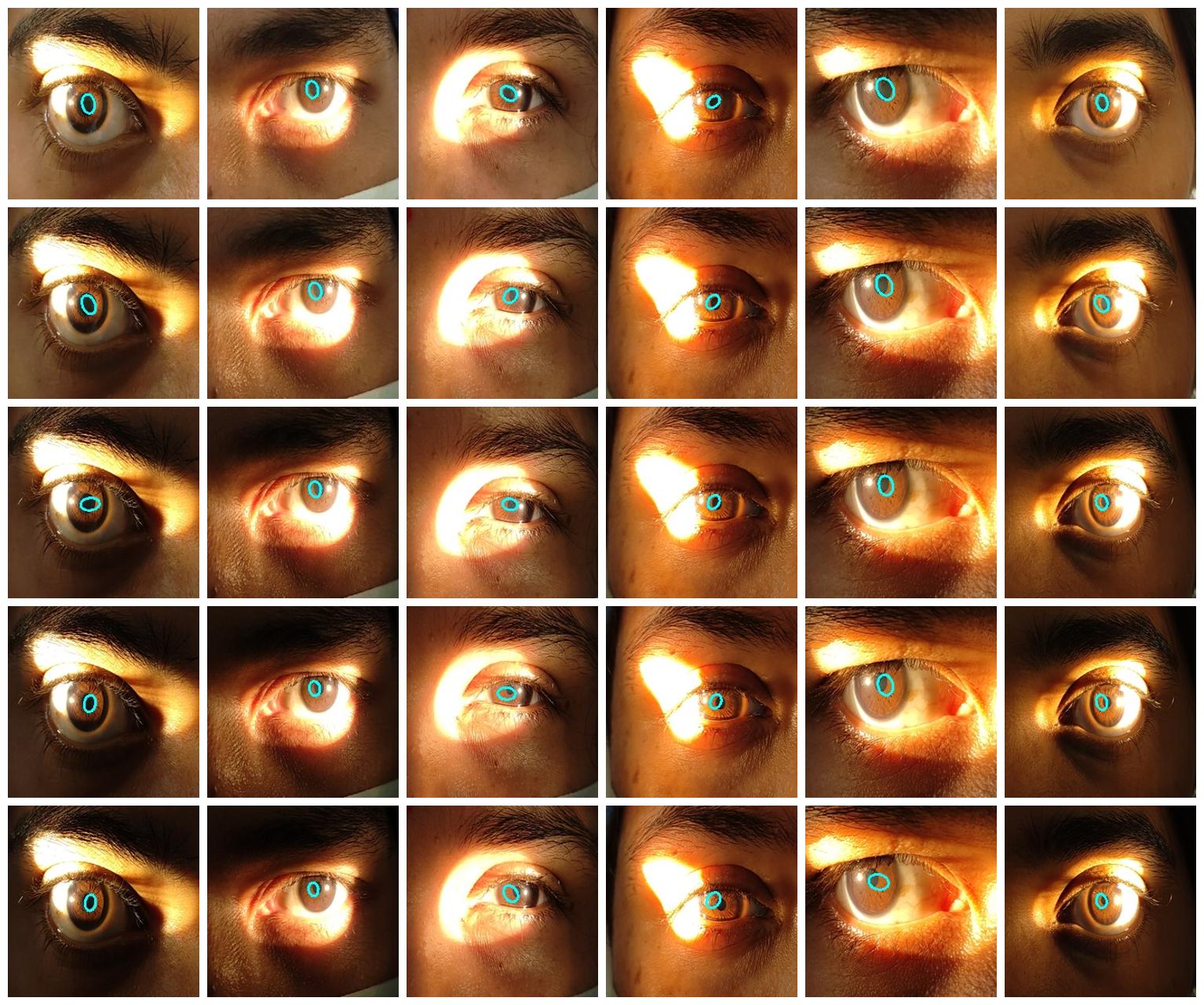} \\
    \small (b)
  \end{tabular}
  \caption{\label{fig:eye5} \redc{Fitted ellipse boundary $\varepsilon_{1}$ directly obtained from U-Net ER model (Method 3) (a) UBIRISv2, (b) Experimentally Acquired Images.}}
\end{figure}

For the experimentally acquired images, we present the results of the neural networks on real-life images which were not used for model training as shown in Fig.~\ref{fig:eye3}, ~\ref{fig:eye4}. We observe that the U-Net ER performs better than the U-Net for real-world data. See columns 3 and 5 of Fig.~\ref{fig:eye3} (c), (d) for some examples. Fitted ellipse boundaries using output ellipse parameters from this model are shown in Fig.~\ref{fig:eye5}.


\redc{
The statistical significance analysis of Accuracy and DSC for tests for the datasets is presented in Table ~\ref{tab:hyptest}. We considered Method 1 as baseline and compared it with Methods 2 and 3 using a two-sided $t$-test with a level of significance $\alpha = 0.05$. For each of the comparisons, we used ($n=30$) samples each for the baseline and the treatment and thus the degrees of freedom ($DoF$) were $58$. We observed that our results were statistically significant ($p < 0.05$) indicating that the observed difference in the means of the methods is due to the method and not due to random effects. 
\begin{table}[ht!]
\caption{Statistical significance analysis} 
\label{tab:hyptest}
\begin{center}       
\begin{tabular}{|c|c|c|c|c|}
\hline
\textbf{Baseline - Treatment} & \textbf{Response} & \textbf{Dataset} & \textbf{$t$-statistic} & \textbf{$p$-value}
\\
\hline
\hline
Method 1 - Method 2 & DSC &  UBIRISv2 &   $-2.537$    &  $0.014$  \\
\hline
Method 1 - Method 2 & Accuracy & UBIRISv2 &   $12.541$   &  $3.68e-18$   \\
\hline
Method 1 - Method 3 & DSC & UBIRISv2 &   $12.308$  &  $8.168e-18$  \\
\hline
Method 1 - Method 3 & Accuracy & UBIRISv2 &   $12.541$   &  $3.68e-18$   \\
\hline
Method 1 - Method 2 &  DSC &  Experimentally Acquired &  $-1.679$  &  $0.01$    \\
\hline
Method 1 - Method 2 & Accuracy &  Experimentally Acquired &  $8.080$ &   $4.498e-11$    \\
\hline 
Method 1 - Method 3 &  DSC &  Experimentally Acquired &   $0.506 $ &  $0.615$    \\
\hline
Method 1 - Method 3 &  Accuracy &  Experimentally Acquired &   $4.532$  &  $2.973e-05$    \\
\hline 
\end{tabular}
\end{center}
\end{table}
}

\section{CONCLUSION}
 A smartphone can potentially be used to perform VL pupillometry. We showed why extraction of ellipse parameters of the pupil is important and how it relates to the measurement of PLR and other important metrics. We demonstrated how Deep Learning can be used to obtain pupil pixels and the ellipse parameters directly in a single step with high accuracy. Training the model with the combined loss function also improved the accuracy and robustness of the model. This can be attributed to the model learning two tasks simultaneously as it learns the correlation between the pupil pixels and the ellipse parameters. Hyper-parameter optimization or the use of different encoder-decoder architectures based on \redc{transformers} for performance or accuracy enhancements can be an interesting direction for the future. \redc{A major limitation of the current work is that we use a training dataset (UBIRISv2) which contain images that differ from the experimentally acquired images. Since these images are samples from a different distribution, previously not seen by the model, the quality of predictions is affected. To make the model predictions more accurate and robust, it is important to collect and label more training data containing samples that are closer to samples in the test set so that the divergence between the two distributions is minimized. However, the collection of VL pupil images and manually annotating the pupil masks is very difficult, time consuming and expensive. Images of the eye (especially irises) are also considered as biometric signatures that can be used for identification thus making such datasets highly sensitive. These issues pertaining to privacy also limit the availability and access to such datasets and present a major challenge for researchers in this area. Applying Generative Models such as Generative Adversarial Networks (GANs) or Diffusion Models can enable the creation of newer (synthetic) examples based on the data distribution and can be advantageous over discriminative modeling.} The work presented here is thus an important step in application of deep learning methods to solve an important task in clinical vision which can be implemented on a smartphone for use under field conditions.

\begin{backmatter}


\bmsection{Disclosures}
The authors declare that there are no conflicts of interest
related to this article.
\bmsection{Data Availability Statement}
Data underlying the results presented in this paper are not publicly available at this time but may be obtained from the authors upon reasonable request.
\end{backmatter}


\bibliography{sample}






\end{document}